\pdfoutput=1
\documentclass[traditabstract]{aa}  

\usepackage{graphicx}
\usepackage{natbib}
\usepackage{morefloats}

\bibpunct{(}{)}{;}{a}{}{,} 
\usepackage{txfonts}
\title{First LOFAR observations at very low frequencies of cluster-scale non-thermal emission: the case of Abell~2256}
\titlerunning{LOFAR observations of cluster-scale non-thermal radio emission in Abell~2256}
   \author{R.~J. van Weeren \inst{\ref{leiden} \and \ref{astron}}
          \and H.~J.~A. R\"ottgering\inst{\ref{leiden}}
          \and D.~A. Rafferty\inst{\ref{leiden}}
          \and R.~Pizzo\inst{\ref{astron}}
          \and A.~Bonafede\inst{\ref{bremen}}
          \and M.~Br\"uggen\inst{\ref{bremen}}
          \and G.~Brunetti\inst{\ref{inaf}}
          \and C.~Ferrari\inst{\ref{nice}}
          \and E.~Orr\`u\inst{\ref{nijmegen}}
          \and G.~Heald\inst{\ref{astron}}
          \and J.~P.~McKean\inst{\ref{astron}}
          \and C.~Tasse\inst{\ref{meudon2}}
          \and F.~de~Gasperin\inst{\ref{mpifa}}
          \and L.~{B{\^i}rzan}\inst{\ref{leiden}}
          \and J.~E. van Zwieten\inst{\ref{astron}}
          \and S.~van~der Tol\inst{\ref{leiden}}
          \and A.~Shulevski\inst{\ref{kapteyn}}
          \and N.~Jackson\inst{\ref{jod}}
          \and A.~R. Offringa\inst{\ref{kapteyn}}
          \and J.~Conway\inst{\ref{oso}}
          \and H.~T. Intema\inst{\ref{nrao}}
          \and T.~E. Clarke\inst{\ref{nrl}}
          \and I.~van~Bemmel\inst{\ref{astron}}
          \and G.~K. Miley\inst{\ref{leiden}}
          \and G.~J. White\inst{\ref{open} \and \ref{ssd}}
          \and M.~Hoeft\inst{\ref{tls}}
          \and R.~Cassano\inst{\ref{inaf}}
          \and G.~Macario\inst{\ref{nice}}
          \and R.~Morganti\inst{\ref{astron} \and \ref{kapteyn}}
          \and M.~W.~Wise\inst{\ref{astron} \and \ref{uva}}
          \and C.~Horellou\inst{\ref{oso}}
          \and E.~A.~Valentijn\inst{\ref{mpifa}}
          \and O.~Wucknitz\inst{\ref{ubonn}}
          \and K.~Kuijken\inst{\ref{leiden}}
          \and T~A.~En{\ss}lin\inst{\ref{mpifa}} 
    	\and J.~Anderson\inst{\ref{mpifr}} \and 
A.~Asgekar\inst{\ref{astron}} \and 
I.~M.~Avruch\inst{\ref{astron} \and \ref{kapteyn}} \and 
R.~Beck\inst{\ref{mpifr}} \and 
M.~E.~Bell\inst{\ref{soton}} \and
M.~R.~Bell\inst{\ref{mpifa}} \and
M.~J.~Bentum\inst{\ref{astron}} \and
G.~Bernardi\inst{\ref{cfa}}\and
P.~Best\inst{\ref{roe}} \and 
A.-J.~Boonstra\inst{\ref{astron}} \and
M.~Brentjens\inst{\ref{astron}} \and 
R.~H.~van de Brink\inst{\ref{astron}} \and 
J.~Broderick\inst{\ref{soton}} \and
W.~N.~Brouw\inst{\ref{astron} \and \ref{kapteyn}}\and
H.~R.~Butcher\inst{\ref{astron} \and \ref{anu}} \and 
W.~van Cappellen\inst{\ref{astron}} \and 
B.~Ciardi\inst{\ref{mpifa}} \and 
J.~Eisl\"offel\inst{\ref{tls}} \and 
H.~Falcke\inst{\ref{nijmegen} \and \ref{astron}} \and 
R.~Fender\inst{\ref{soton}}\and
M.~A.~Garrett\inst{\ref{astron} \and \ref{leiden}} \and 
M.~Gerbers\inst{\ref{astron}} \and 
A.~Gunst\inst{\ref{astron}} \and 
J.~P.~Hamaker\inst{\ref{astron}}
T.~Hassall\inst{\ref{jod}}\and
J.~W.~T.~Hessels\inst{\ref{astron} \and \ref{uva}} \and
L.~V.~E.~Koopmans\inst{\ref{kapteyn}} \and 
G.~Kuper\inst{\ref{astron}}\and
J.~van Leeuwen\inst{\ref{astron} \and \ref{uva}}\and
P.~Maat\inst{\ref{astron}} \and 
R.~Millenaar\inst{\ref{astron}} \and 
H.~Munk\inst{\ref{astron}} \and
R.~Nijboer\inst{\ref{astron}} \and 
J.~E.~Noordam\inst{\ref{astron}} \and 
V.~N.~Pandey\inst{\ref{kapteyn}} \and 
M.~Pandey-Pommier\inst{\ref{leiden} \and \ref{lyon}} \and 
A.~Polatidis\inst{\ref{astron}} \and 
W.~Reich\inst{\ref{mpifr}} \and 
A.~M.~M.~Scaife\inst{\ref{soton}} \and 
A.~Schoenmakers\inst{\ref{astron}} \and 
J.~Sluman\inst{\ref{astron}} \and 
B.~W.~Stappers\inst{\ref{jod}} \and 
M.~Steinmetz\inst{\ref{aip}} \and 
J.~Swinbank\inst{\ref{uva}} \and 
M.~Tagger\inst{\ref{cnrs}} \and 
Y.~Tang\inst{\ref{astron}} \and 
R.~Vermeulen\inst{\ref{astron}} \and 
M.~de Vos\inst{\ref{astron}}
}

\institute{Leiden Observatory, Leiden University, PO Box 9513, 2300 RA Leiden, The Netherlands\label{leiden}\\
\email{rvweeren@strw.leidenuniv.nl}
\and Netherlands Institute for Radio Astronomy (ASTRON), Postbus 2, 7990 AA Dwingeloo, The Netherlands\label{astron}
\and Jacobs University Bremen, Campus Ring 1, 28759 Bremen, Germany\label{bremen}
\and INAF/Istituto di Radioastronomia, via Gobetti 101, I-40129 Bologna, Italy\label{inaf}
\and Laboratoire Lagrange, UMR7293, Universit\'e de Nice Sophia-Antipolis, CNRS, Observatoire de la C\^ote d'Azur, 06300 Nice, France\label{nice}
\and Department of Astrophysics/IMAPP, Radboud University Nijmegen, P.O. Box 9010, 6500 GL Nijmegen, The Netherlands\label{nijmegen}
\and GEPI, Observatoire de Paris-Meudon, 5 place Jules Janssen, 92190 Meudon, France\label{meudon2}
\and Max Planck Institute for Astrophysics, Karl Schwarzschild Str. 1, 85741 Garching, Germany\label{mpifa}
\and Kapteyn Astronomical Institute, PO Box 800, 9700 AV Groningen, The Netherlands\label{kapteyn}
\and Jodrell Bank Center for Astrophysics, School of Physics and Astronomy, The University of Manchester, Manchester M13 9PL, UK\label{jod}
\and Onsala Space Observatory, Dept. of Earth and Space Sciences, Chalmers University of Technology, SE-43992 Onsala, Sweden\label{oso}
\and National Radio Astronomy Observatory, 520 Edgemont Road, Charlottesville, VA 22903-2475, USA\label{nrao}
\and Naval Research Laboratory, 4555 Overlook Avenue SW, Washington~D.~C. 20375, USA\label{nrl}
\and Department of Physics \& Astronomy, The Open University, UK\label{open}
\and Space Science Department, Rutherford Appleton Laboratory, Chilton, UK\label{ssd}
\and Th\"uringer Landessternwarte, Sternwarte 5, D-07778 Tautenburg, Germany\label{tls}
\and Astronomical Institute `Anton Pannekoek', University of Amsterdam, Postbus 94249, 1090 GE Amsterdam, The Netherlands\label{uva}
\and Argelander-Institut f\"ur Astronomie, University of Bonn, Auf dem H\"ugel 71, 53121, Bonn, Germany\label{ubonn}
\and Max-Planck-Institut f\"ur Radioastronomie, Auf dem H\"ugel 69, 53121 Bonn, Germany\label{mpifr}
\and School of Physics and Astronomy, University of Southampton, Southampton, SO17 1BJ, UK\label{soton}
\and Harvard-Smithsonian Center for Astrophysics, Garden Street 60, Cambridge, MA, 02138, USA\label{cfa}
\and Institute for Astronomy, University of Edinburgh, Royal Observatory of Edinburgh, Blackford Hill, Edinburgh EH9 3HJ, UK\label{roe}
\and Mt Stromlo Obs., Research School of Astronomy and Astrophysics, Australian National University, Weston, A.C.T. 2611, Australia\label{anu}
\and Centre de Recherche Astrophysique de Lyon, Observatoire de Lyon, 9 av Charles Andr\'e, 69561 Saint Genis Laval Cedex, France\label{lyon}
\and Leibniz-Institut f\"ur Astrophysik Potsdam (AIP), An der Sternwarte 16, 14482 Potsdam, Germany\label{aip}
\and Laboratoire de Physique et Chimie de l'Environnement et de l'Espace 3A, Avenue de la Recherche Scientifique 45071 Orleans cedex 2, France\label{cnrs}
}

   \date{}

\begin{document}
\abstract
    {Abell~2256 is one of the best known examples of a galaxy cluster hosting large-scale diffuse radio emission that is unrelated to individual galaxies. It contains both a giant radio halo and a relic,  as well as a number of head-tail sources and smaller diffuse steep-spectrum radio sources. The origin of radio halos and relics is still being debated, but over the last years it has become clear that the presence of these radio sources is closely related to galaxy cluster merger events. Here we present the results from the first LOFAR Low band antenna (LBA) observations of Abell~2256 between 18 and 67~MHz. To our knowledge, the image presented in this paper at  63~MHz is the deepest ever obtained at frequencies below 100~MHz in general. 
     Both the radio halo and the giant relic are detected in the image at 63~MHz, and the diffuse radio emission remains visible at frequencies as low as 20 MHz.  
    The observations confirm the presence of a previously claimed ultra-steep spectrum source to the west of the cluster center with a spectral index of $-2.3 \pm 0.4$ between 63 and 153~MHz. The steep spectrum suggests that this source is an old part of a head-tail radio source in the cluster. For the radio relic we find an integrated spectral index of $-0.81 \pm 0.03$, after removing the flux contribution from the other sources. This is relatively flat which could indicate that the efficiency of particle acceleration at the shock  substantially changed in the last $\sim0.1$~Gyr due to an increase of the shock Mach number.  
    In an alternative scenario, particles are re-accelerated by some mechanism in the downstream region of the shock, resulting in the relatively flat integrated radio spectrum. In the radio halo region we find indications of low-frequency spectral steepening which may suggest that relativistic particles are accelerated in a rather inhomogeneous turbulent region.  
 }

      \maketitle

\section{Introduction}

Radio halos and relics are diffuse radio sources, unrelated to individual galaxies, found in some disturbed galaxy clusters \citep[see the reviews by][and references therein]{2005AdSpR..36..729F, 2008SSRv..134...93F, 2011MmSAI..82..499V}. These sources are not ubiquitous in galaxy clusters  -- only a few dozen examples have been previously identified. The presence of diffuse radio emission indicates magnetic fields and relativistic particles in the intracluster medium (ICM). Due to their limited lifetime, the synchrotron emitting electrons need to be (re)accelerated in-situ \citep{1977ApJ...212....1J}. The origin of this diffuse emission is presently still being debated and is crucial for  understanding the non-thermal component in the ICM and particle acceleration mechanisms.

It is predicted in the framework of the concordant cosmological model that galaxy clusters grow as a result of the mergers of smaller clusters and sub-structures, which create turbulence and shocks in their ICM \citep[e.g.,][]{2002ApJ...567L..27M, 2006MNRAS.369L..14V, 2010MNRAS.406.1721R}. Over the last decade it has become clear that the diffuse radio emission in galaxy clusters is often related to galaxy cluster merger events \citep[e.g.,][]{2010ApJ...721L..82C, 2010Sci...330..347V}.

Diffuse sources in merging clusters have commonly been classified as radio {\it relics} and {\it halos}. In this context, elongated filamentary sources have generally been called relics. Relics are usually further subdivided into three classes \citep[see also][]{2004rcfg.proc..335K}. (1) {\it Radio Gischt} are large extended arc-like sources mostly found in the outskirts of galaxy clusters. They are often highly polarized, 20\% or more, at frequencies $\gtrsim 1$~GHz.
It has been proposed that they directly trace shock waves \citep{1998A&A...332..395E, 2001ApJ...562..233M}, in which particles are accelerated by the diffusive shock acceleration mechanism in a first-order Fermi process \citep[DSA;][]{1978ApJ...221L..29B, 1983RPPh...46..973D, 1987PhR...154....1B, 1991SSRv...58..259J, 2001RPPh...64..429M}. A related scenario is that of shock re-acceleration of pre-accelerated electrons in the ICM. This may be a more efficient mechanism in weak shocks  \citep[e.g.,][]{2005ApJ...627..733M, 2008A&A...486..347G, 2011ApJ...734...18K} and might be needed because the efficiency with which collisionless shocks can accelerate particles is unknown, and might not be enough to produce the observed radio brightness \citep[e.g.,][]{2011ApJ...728...82M}. The pre-accelerated electrons could for example originate from the radio galaxies in clusters. 
 (2) {\it Radio phoenices} are most likely related to fossil radio plasma from previous episodes of AGN activity adiabatically compressed by merger shock waves, boosting the radio emission \citep{2001A&A...366...26E, 2002MNRAS.331.1011E}. (3) {\it AGN relics} trace uncompressed radio plasma from previous episodes of AGN activity. The radio spectra of both radio phoenices and AGN relics are expected to be steep\footnote{$\alpha \lesssim -1.5$, $F_{\nu} \propto \nu^{\alpha}$, where $\alpha$ is the spectral index} and curved  due to synchrotron and Inverse Compton (IC) losses.

\emph{Radio halos} are extended  sources with sizes of about a Mpc. They are usually unpolarized, with limits at the several percent level, and the radio emission roughly follows the thermal X-rays from the ICM, implying that the emitting relativistic plasma is co-spatial with the thermal gas in galaxy clusters.
Potentially several mechanisms may be responsible for the 
generation of Mpc scale synchrotron emission from galaxy clusters. 
Relativistic electrons in the ICM can be re-accelerated
in-situ through interaction with turbulence generated in the ICM by cluster-cluster 
mergers \citep[][]{2001MNRAS.320..365B, 2001ApJ...557..560P}, or 
secondary electrons can be continuously injected in the ICM 
by inelastic collisions between
relativistic and thermal protons  \citep[][]{1980ApJ...239L..93D, 1999APh....12..169B, 2000A&A...362..151D, 2011A&A...527A..99E}.
Models considering a combination of the two mechanisms, namely the
re-acceleration of both relativistic protons and their secondaries by magnetohydrodynamical turbulence
generated in the ICM have been recently considered \citep{2005MNRAS.363.1173B,2008SSRv..134..311D,2011MNRAS.410..127B}.

In the case of several giant radio halos, a pure secondary origin is disfavored
from their spectra and/or large extension \citep[e.g.,][]{2008Natur.455..944B,2009ApJ...699.1288D,2010MNRAS.401...47D,2011MNRAS.412....2B}. 
In particular, an unrealistic amount of energy in the form of relativistic protons 
would be required by secondary models to explain radio halos with very steep
spectra \citep{2004JKAS...37..493B,2004MNRAS.352...76P,2008Natur.455..944B,2010A&A...517A..43M,2011A&A...533A..35V}. 
On the other hand, if turbulence plays an important role for the origin of these 
giant sources, a large number of steep spectrum radio halos is expected to glow up
at low radio frequencies \citep{2006MNRAS.369.1577C,2008Natur.455..944B}. The origin of the turbulent magnetic field remains unclear. One possibility is that it has been amplified by dynamo processes acting in the cluster volume \citep[e.g.,][and references therein]{1989MNRAS.241....1R,2005JCAP...01..009D}.

The LOw Frequency ARray (LOFAR) is a new generation radio telescope operating at 10--240~MHz (van Haarlem et al., in prep). The large collecting area and instantaneous fractional bandwidth, together with long baselines, allow the first detailed study of low-surface brightness diffuse cluster radio sources below 100 MHz. In this respect LOFAR is the ideal radio telescope to unveil the existence of steep spectrum diffuse radio emission from galaxy clusters and is expected to
provide a breakthrough in the field \citep{2010A&A...509A..68C}.

\object{Abell~2256} is a nearby, $z=0.0581$ \citep{1999ApJS..125...35S},  cluster that contains a giant 1.2~Mpc radio halo, a relic, and a number of tailed radio galaxies \citep{1976A&A....52..107B, 1979A&A....80..201B, 1994ApJ...436..654R, 2003AJ....125.2393M, intema_phd, 2009A&A...508.1269V, 2010ApJ...718..939K}. The relic has a large integrated flux, compared to other relics, of about 0.5~Jy at 1.4~GHz, and a size of 1125~kpc~$\times$~520~kpc. The average polarization fraction of the relic is about 20\% at 1.4~GHz. \cite{2006AJ....131.2900C}. A spectral analysis by \cite{2008A&A...489...69B} shows that the radio halo component dominates the integrated cluster radio spectrum at very low frequencies. X-ray and optical observations provide strong evidence that A2256 is undergoing a merger event between a main cluster ($T_{\rm{ICM}} \sim 7$--8~keV), a major sub-structure ($T_{\rm{ICM}} \sim 4.5$ keV) and, possibly, a third infalling group \citep{1991A&A...246L..10B, 1994Natur.372..439B,2002ApJ...565..867S, 2002AJ....123.2261B,2003AJ....125.2393M}. The cluster has an X-ray luminosity of $L_{\rm{X,~0.1-2.4~keV}} = 3.7 \times 10^{44}$~erg~s$^{-1}$ \citep{1998MNRAS.301..881E}.

\begin{figure*}
   \begin{center}
       \includegraphics[angle = 0, trim =0.6cm 0.6cm 0.6cm 0.6cm,width=0.49\textwidth]{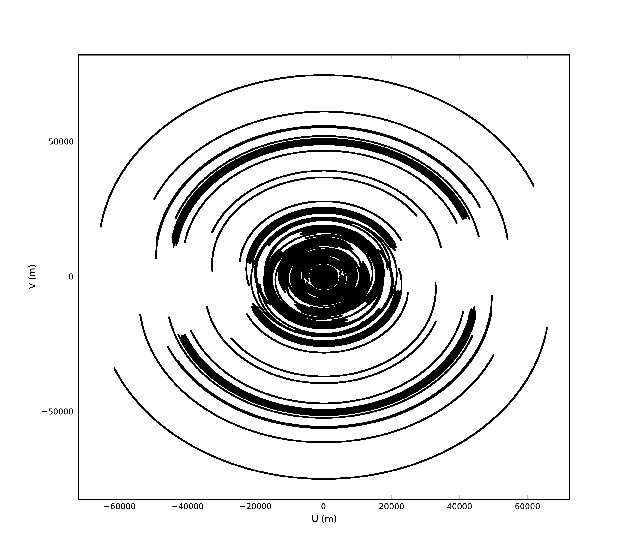}
       \includegraphics[angle = 0, trim =0.6cm 0.6cm 0.6cm 0.6cm,width=0.49\textwidth]{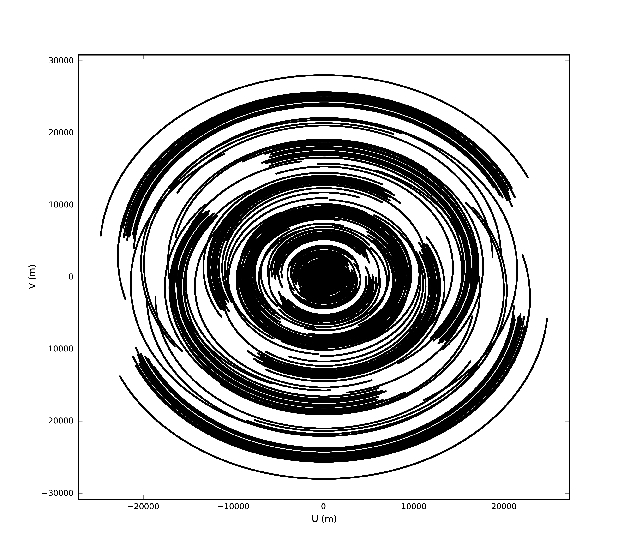}
    \includegraphics[angle = 0, trim =0.6cm 0.6cm 0.6cm 0.6cm,width=0.49\textwidth]{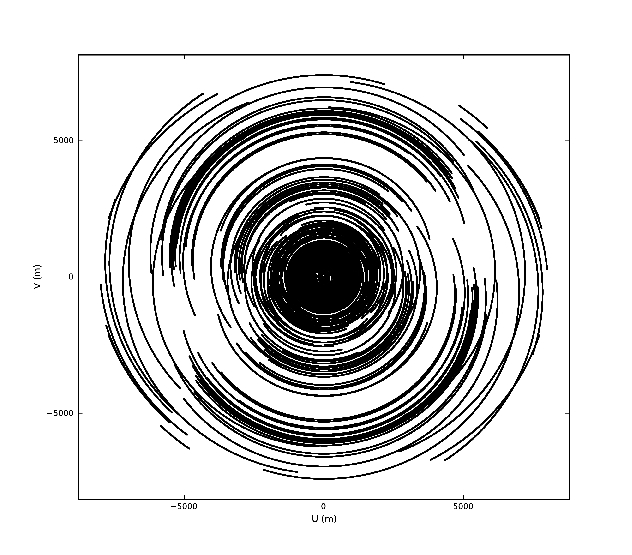}
    \includegraphics[angle = 0, trim =0.6cm 0.6cm 0.6cm 0.6cm,width=0.49\textwidth]{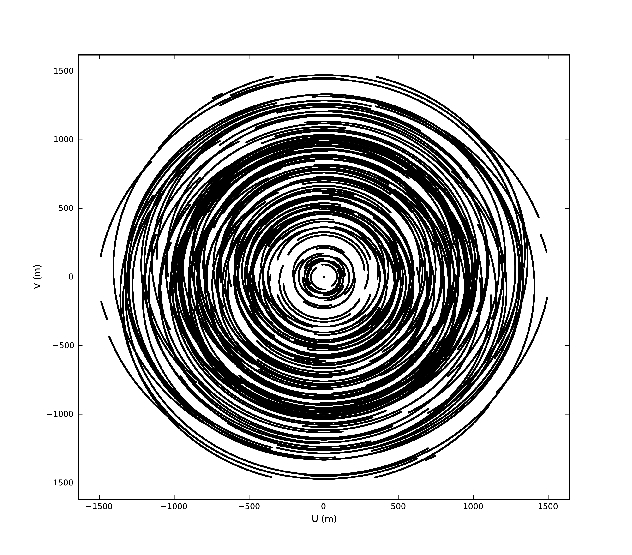}
     \end{center}
    \caption{UV-coverage of the A2256 observation on July 15 and 16, 2011, from 18:00 to 04:00 CEST (10 hrs in total). Transit occurred around 23:00 CEST.     
    The outer UV-coverage is shown in the top left frame, the next frames progressively zoom inwards. The relatively broad bandwidth fills the uv-plane radially (not shown in the figures). }
    \label{fig:lbauv}
\end{figure*}

The large angular extent of the diffuse emission and its large integrated flux make A2256 a prime target for low-frequency observations which typically suffer from low spatial resolution and sensitivity, compared to observations at high frequencies. At wavelengths larger than 1~m, the diffuse radio emission in A2256 has been studied with the GMRT at 150~MHz \citep{intema_phd, 2010ApJ...718..939K}, and with the WSRT between 115 and 165 MHz \citep{2009A&A...508.1269V, intema_phd}. The VLSS 74~MHz image \citep{2007AJ....134.1245C} only shows a hint of diffuse emission, and the 38~MHz 8C and 151~MHz 6C survey images \citep{1995MNRAS.274..447H, 1978MNRAS.185..607M} have too low resolution to properly separate the flux contributions from the discrete sources. The observations by \cite{intema_phd} and \cite{2009A&A...508.1269V} showed the presence of three previously unknown diffuse sources with steep radio spectra ($\alpha \lesssim -1.5$). These steep radio spectra suggest that these sources trace old radio plasma from previous episodes of AGN activity in the cluster.

 In this paper we present the first LOFAR  observations of Abell~2256, focussing on the frequency range around 63~MHz and the total intensity (stokes I) images. The layout of this paper is as follows: in Sect.~\ref{sec:obs-reduction} we give an overview of the observations and the data reduction. In Sect.~\ref{sec:results} we present the results, including the radio images and a spectral index map. We end with a discussion and the conclusions in  Sects.~\ref{sec:discussion} and \ref{sec:conclusion}, respectively.
Throughout this paper we assume a $\Lambda$CDM cosmology with $H_{0} = 71$~km~s$^{-1}$~Mpc$^{-1}$, $\Omega_{\rm{m}} = 0.3$, and $\Omega_{\Lambda} = 0.7$.

\section{Observations \& data reduction}
\label{sec:obs-reduction}

\subsection{Observations}

\begin{figure}
   \begin{center}
   \includegraphics[angle = 0, trim =0cm 0cm 0cm 0cm,width=0.51\textwidth]{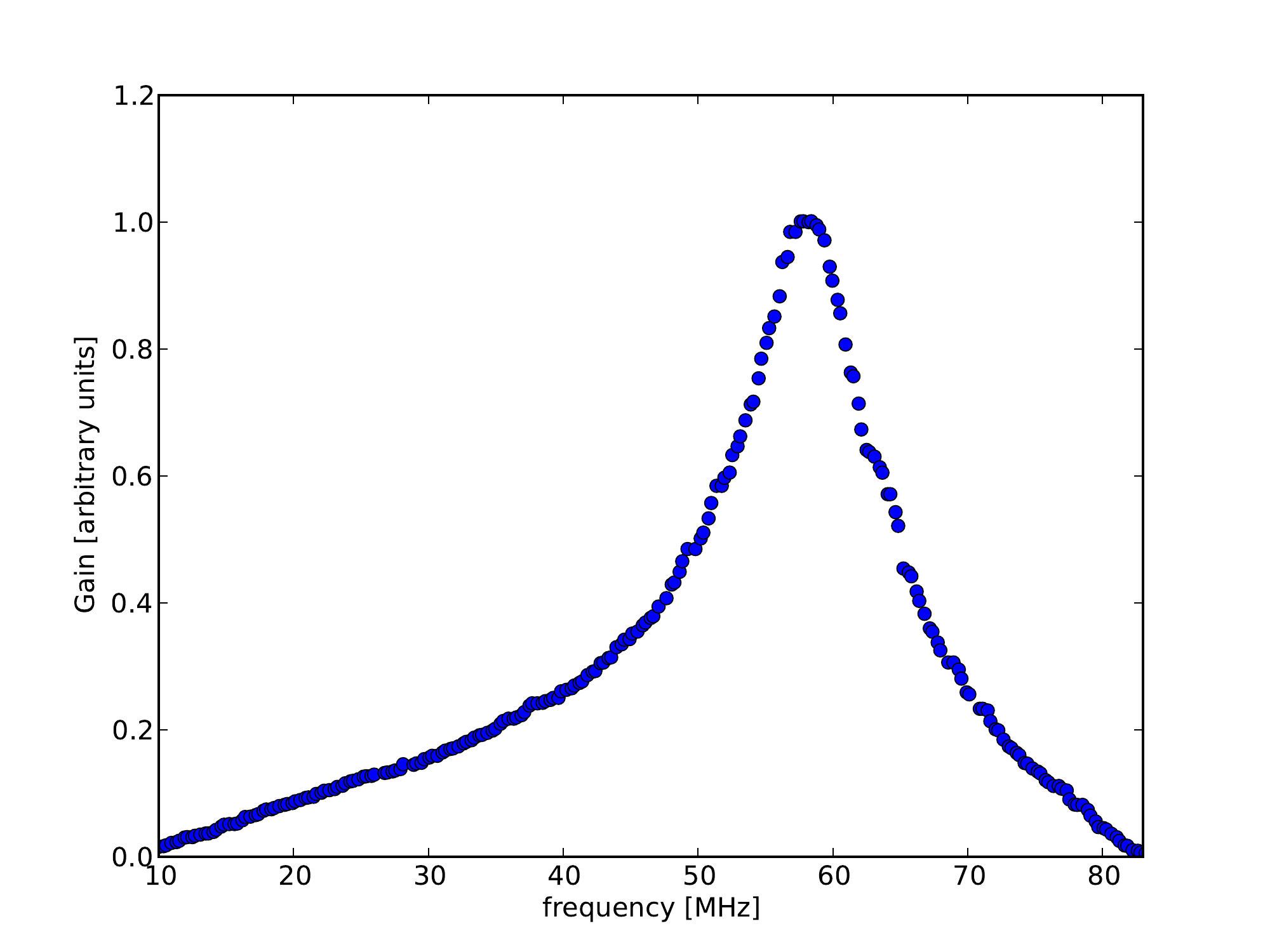}
     \end{center}
    \caption{LBA global bandpass. The bandpass peaks at about 58~MHz as expected due to resonance.}
    \label{fig:LBAbandpass}
\end{figure}

\begin{figure}
   \begin{center}
   \includegraphics[angle = 90, trim =0cm 0cm 0cm 0cm,width=0.47\textwidth]{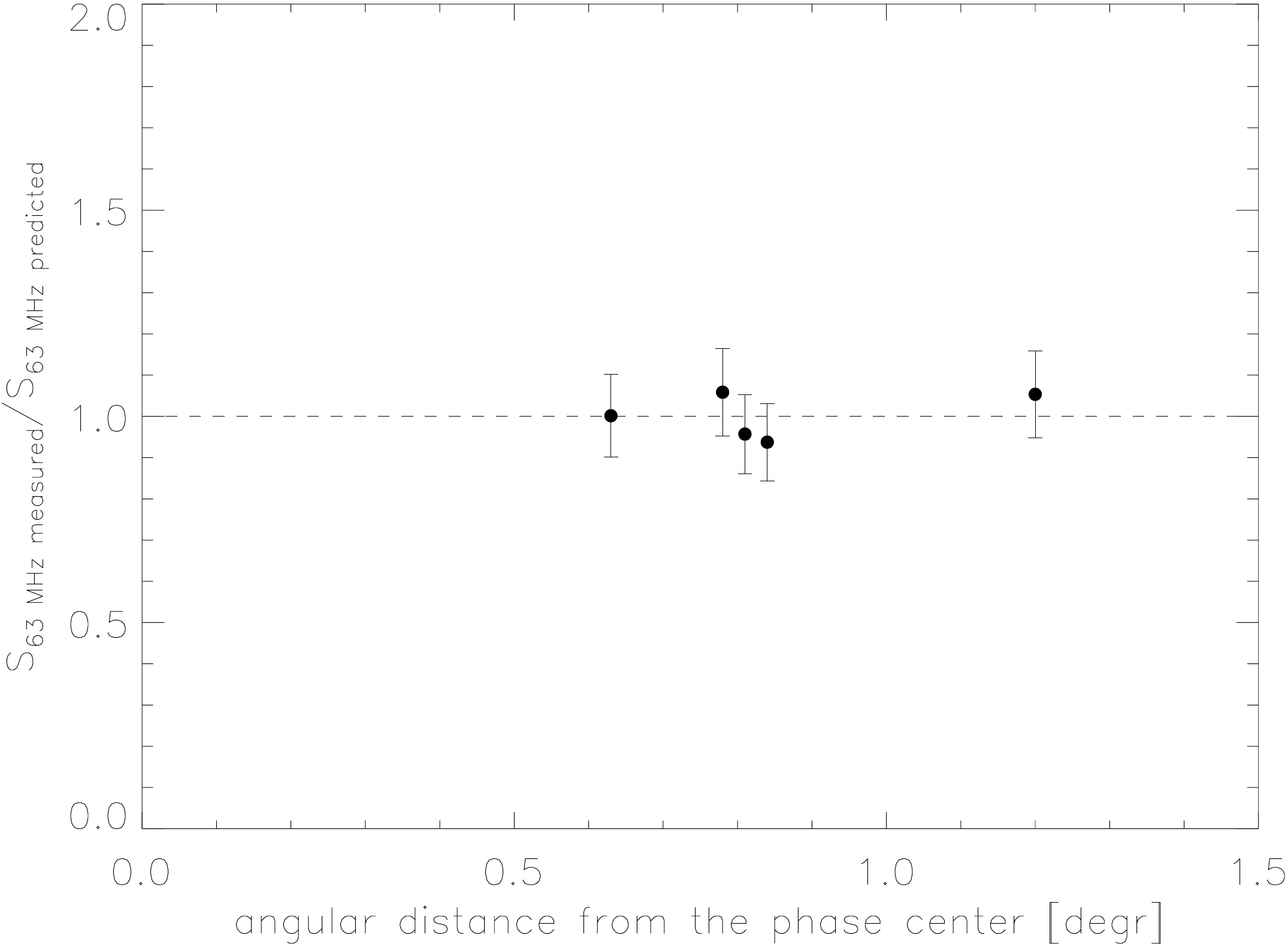}
     \end{center}
    \caption{Measured LOFAR fluxes and predicted fluxes for 5 bright sources in the A2256 field. For each of these sources, 8C, VLSS, 7C, WENSS, and NVSS flux densities were available. We used second order polynomial fits in  $\log{(S)}-\log{(\nu)}$ space to compute the predicted 63~MHz fluxes. An overall scaling factor of 1.23 was applied to the LOFAR fluxes. The error bars display the adopted 10\% uncertainty in the flux scale.}
    \label{fig:LBAfluxratio}
\end{figure}

Abell 2256~was observed with the LOFAR LBA system on July 15 and 16, 2011, from 18:00 to 04:00 CEST. 
The observations covered the 18 to 67~MHz frequency range, except for a gap at 40--42~MHz due to strong radio frequency interference (RFI). All four linear correlation products were recorded. The observed frequency range was divided into 244 sub-bands, each having a bandwidth of 0.1953~MHz. Each sub-band was subdivided into 64 frequency channels. The integration time was 1~s. The so-called {LBA\_OUTER} configuration was used for the LBA stations, where LBA antenna refers to a dipole pair and station refers to the collection of dipole pairs grouped as one. In the {LBA\_OUTER} configuration 48 (of 96) LBA antennas are used, located mostly in the outer part of the stations (which have diameters of 87~m). This increases the sidelobe levels for the station beams but reduces the field of view (FOV) with respect to other station antenna configurations.

We used eight remote and seventeen core stations for the A2256 observation. The baselines are between 90 m and 80 km long. 
The uv-coverage is shown in Fig.~\ref{fig:lbauv}. The FOV (full width half maximum (FWHM) of the primary beam) is about 4.6\degr~and 9.2\degr~at 60 and 30~MHz, respectively. It  should be noted though that the station beams are complex-valued, time and direction dependent, and differ slightly from station to station.

As a first step, we flagged RFI with the AOFlagger \citep{2010MNRAS.405..155O} using default settings. The first and last two channels at the edge of each sub-band were removed. 
The flagger typically found up to a few percent of RFI above 30 MHz. The amount of RFI increases strongly below $\sim28$~MHz. Between 30\% and 70\% had to be flagged below this frequency for most sub-bands, with at least some of this attributable to short-wave radio transmissions. The RFI situation was somewhat better after midnight. After flagging, we averaged the data in time to 5~s, see \cite{2010iska.meetE..57H} for a brief overview of the data processing steps. 

\subsection{Calibration}

For this A2256 observation, the bright ``A-team'' radio sources Cas~A, Cyg~A and Vir~A are located 34\degr,  42\degr, and 74\degr~away from the phase center, respectively.  These sources have integrated flux densities of about 18, 17, and 2.1~kJy at 74~MHz, respectively.  Even after attenuation by the primary beam, Cas A and Cyg A outshine the brightest sources in the target field by more than a factor ten. Their amplitudes are strongly modulated as they move in and out of the station beam sidelobes. For the first 2.5~hr of the observations Vir~A also affects the observed visibilities, until the source sets. At frequencies  $\lesssim 35$~MHz, 3C~390.3 (located 4.7\degr~from the phase center) is sufficiently bright, $145 \pm 14$~Jy at 38~MHz \citep{1995MNRAS.274..447H}, that it needs to be treated separately as described below.

The first calibration step consisted of the removal of the three ``A-team'' sources. Below 35~MHz, 3C~390.3 was also included in the calibration model. For the models of the A-team sources and 3C~390.3 we used the clean component models at 74~MHz from VLA A-array\footnote{http://lwa.nrl.navy.mil/tutorial/} observations \citep{2007ApJS..172..686K} with a resolution of 25\arcsec. 
For all the calibration steps we used the BlackBoard Selfcal (BBS) software system \citep{2009ASPC..407..384P}. Full polarization direction dependent complex gain solutions were obtained for the three A-team sources (and 3C~390.3 below 35~MHz) independently per sub-band. Vir~A was only included in the calibration model for the first 2.5~hr.
In general, we obtained high-quality gain solutions for Cas~A and Cyg~A, while the solutions for Vir~A and 3C~390.3 were noisier due to their lower apparent fluxes. The A-team sources were then subtracted (like ``peeling'')  using their direction dependent gain solutions from the visibility data. 3C390.3 was not subtracted because it is in the main FOV below 35~MHz and used in subsequent calibration steps.

After subtraction of the A-team sources from the visibility data we performed another round of flagging with the AOFlagger and averaged the data to 4 channels and 10 sec per sub-band, to decrease the size of the dataset whilst limiting	bandwidth and time smearing in the field of view to acceptable levels.

The responses of the LBA antennas depend on the observed frequency, see Fig.~\ref{fig:LBAbandpass}. The LBA response was obtained from  observations of Cyg~A.  Observations that have been carried out over the course of several months show that the bandpass response is stable at the level of a few percent or better. We divided out this sensitivity pattern to avoid the need to obtain amplitude calibration solutions for each individual sub-band. The reason behind this is that the signal to noise ratio (SNR) per sub-band is not sufficient to obtain good gain solutions due to the limited effective bandwidth of 0.183~MHz. By combining several sub-bands, we can overcome this limitation. However, a problem at low frequencies is that the ionospheric phase distortions are frequency dependent, increasing towards lower frequencies. At about 60~MHz this means that if this effect is not included only about 1~MHz bandwidth can be used for calibration, depending on the ionospheric conditions \citep[e.g., Eq.~4 from][]{2009A&A...501.1185I}.
 To obtain high-quality solutions  more bandwidth is required. This can be accomplished by solving for the differential total electron content (TEC).  
 This adds a frequency dependent phase term ($\rm{TEC}/\nu$). During the calibration we therefore solve for (i) a single overall polarization dependent complex gain factor to set the amplitudes and capture other instrumental effects such as clock drift, and (ii) a polarization independent TEC value per station and time-slot (10~s). We used 20 sub-bands around 20~MHz, 20 around 30~MHz, and 30 around 63~MHz per solution interval for this ``global calibration''.  
 Due to computational limitations, we concentrated our efforts on these three frequency ranges, using in total only $70$ of the 244  subbands observed.   
 
We calibrated the data against a 74~MHz 80\arcsec~VLSS model of the field around A2256, covering $15\degr\times15\degr$, and assumed that all sources are unpolarized. This is a reasonable assumption at these low frequencies. We used an overall flux scaling with a spectral index of $-0.8$ to get a first order approximation of the flux-scale. For computing the model visibilities, we included the complex beam attenuation of each station beam. The station beam model is derived using the dipole beam model based on interpolation of electromagnetic simulations of LBA dipole beam response. The interpolation scheme is described by \cite{hamaker}. The beam model does not take into account mutual coupling effects. It is also been assumed that all the dipole pairs/antennas have identical beams.

\begin{figure*}
   \begin{center}
   \includegraphics[angle = 90, trim =0cm 0cm 0cm 0cm,width=0.46\textwidth]{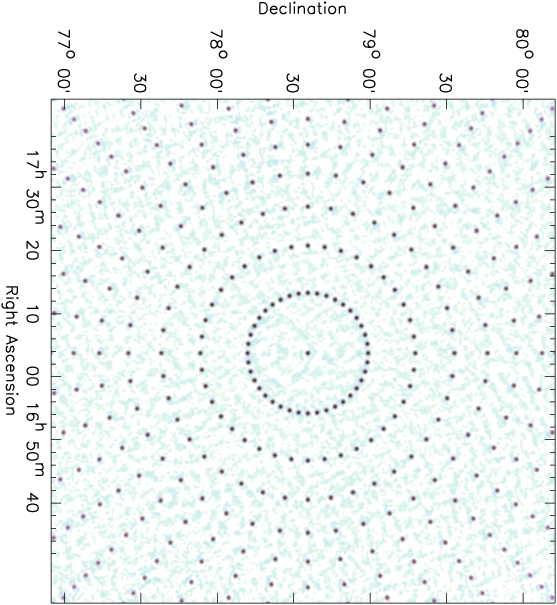}
   \includegraphics[angle = 90, trim =0cm 0cm 0cm 0cm,width=0.53\textwidth]{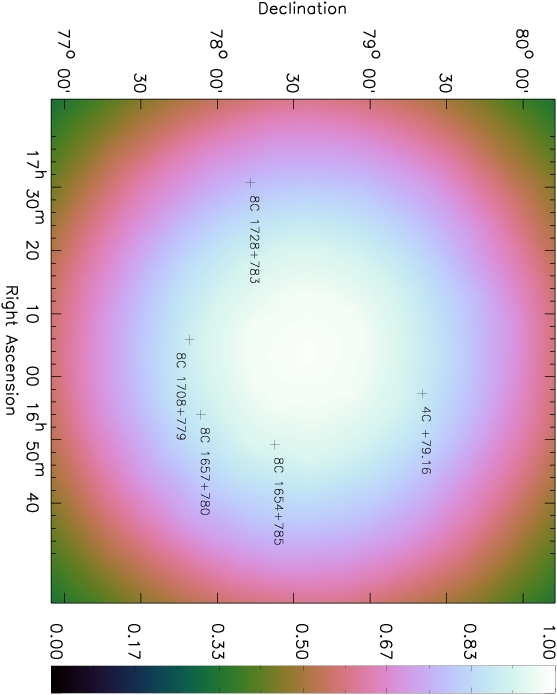}
     \end{center}
    \caption{{\it Left}: Image of the simulated 1~Jy point sources in the A2256 field at 63~MHz. Note that the background is not precisely zero  because of deconvolution errors. The apparent flux of the sources decreases radially away from the phase center due to the primary beam attenuation. Color-scale is the same as in the right panel but inverted. {\it Right}: Primary beam attenuation of the A2256 observations at 63MHz obtained by fitting a surface to the integrated fluxes of the sources in the left panel. The location of sources that were used to obtain the flux-scale are marked with crosses.}
    \label{fig:LBApb}
\end{figure*}

\begin{figure}
   \begin{center}
   \includegraphics[angle = 0, trim =0cm 0cm 0cm 0cm,width=0.5\textwidth]{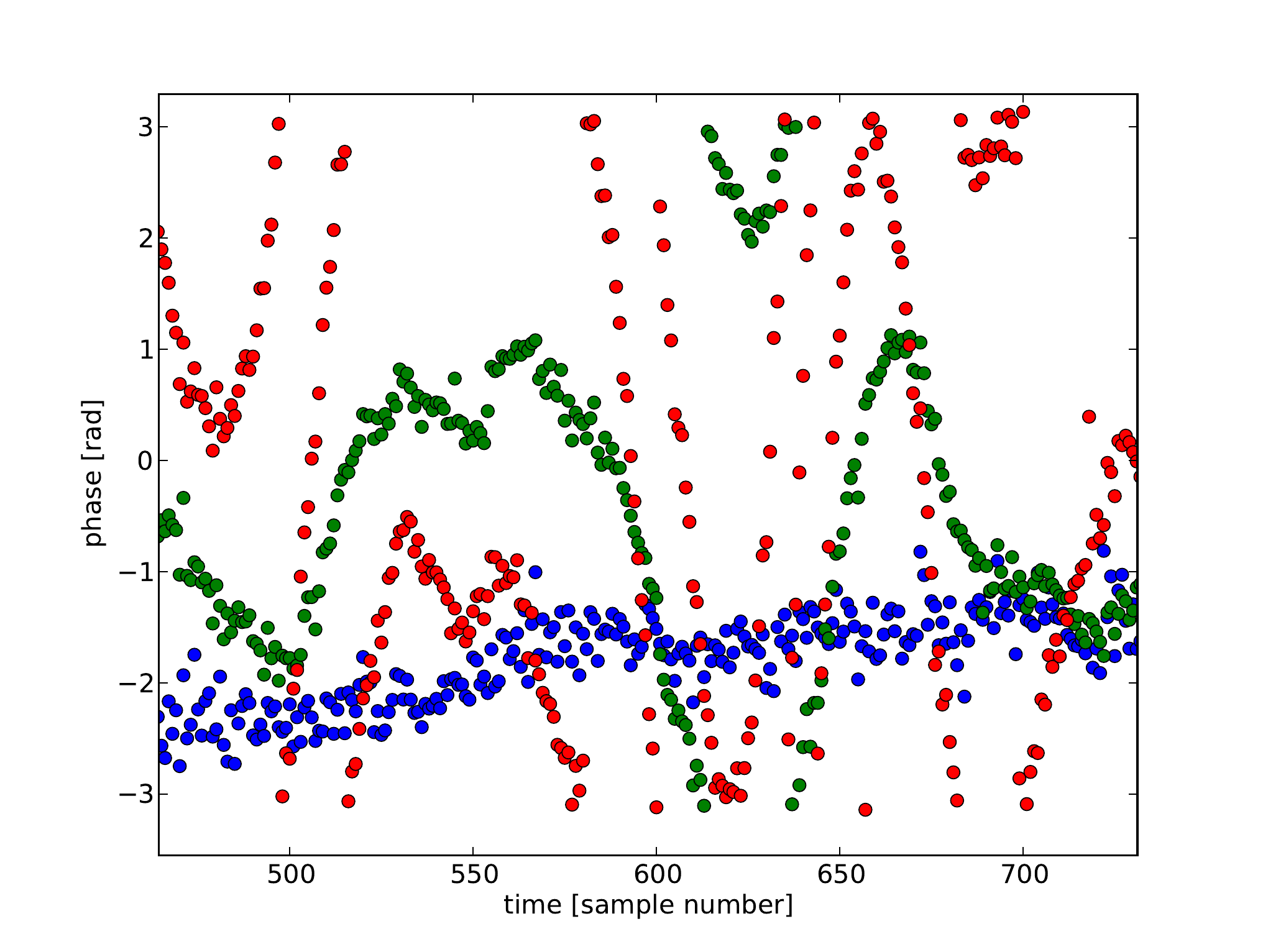}
     \end{center}
    \caption{Phase self-calibration solutions for stations CS017 (blue), RS205 (green), RS208 (red) at 63.9~MHz. Each time sample corresponds to 10~s and the phases are plotted relative to station CS003. The distances between the stations CS017, RS205, RS208 and core station CS003 are about 0.5, 7, and 29~km, respectively.}
    \label{fig:LBAphasesol}
\end{figure}

\subsection{Primary beam correction, flux-scale, and self-calibration}
We applied the calibration solutions and corrected the data for each station's beam response in the phase center. Note that it is only possible to correct for a stationÕs beam response in a single direction. For the other directions within the main FOV this correction is only a first order approximation. 
The entire FOV, to an attenuation factor of $0.15$ compared to the beam center, was then imaged and cleaned using Casapy with w-projection \citep{2005ASPC..347...86C,2008ISTSP...2..647C}; 768 w-planes were used in total.  Ideally, to create flux corrected wide-field images, the time-variable direction dependent effects need to be taken into account \citep[e.g.,][]{2008A&A...487..419B}. This  functionality is still under development for LOFAR data. 
We made images with ``briggs'' weighting \citep{briggs_phd} with robust~=~$-0.1$. This gives a good resolution of around 25\arcsec at 63 MHz, while it still reveals some of the diffuse emission. Using lower values of the robust parameter is not very useful for this dataset because the effective resolution is limited by the ionosphere (as explained later in this subsection). We combined groups of sub-bands around 63, 30 and 20~MHz for multi-frequency synthesis imaging, producing three broadband images.

We simulated a dense grid of 1~Jy point sources in the A2256 field, including the LOFAR station beam, to obtain the correct relative flux scale over the field. This grid is illustrated in Fig.~\ref{fig:LBApb}. We imaged this simulated dataset (with the same frequency setup and uv-coverage as the observed data) in the same way as the observed A2256 data. We extracted the integrated fluxes for the simulated point sources using the {\tt PyBDSM} source detection software\footnote{see the LOFAR Imaging Cookbook at \newline http://www.astron.nl/radio-observatory/lofar/lofar-imaging-cookbook and http://www.strw.leidenuniv.nl/\~{ }mohan/anaamika manual.pdf}. A 2D surface was fitted to these extracted source fluxes using the {\tt griddata} module from the python matplotlib, which employs a Delaunay triangulation\footnote{http://matplotlib.sourceforge.net}. This triangulated surface gives the effective sensitivity as a function of position in the A2256 field and can be used to create a primary beam corrected image. This procedure for correcting a uniform flux scale within the primary beam is a useful stop-gap solution until an a truly beam-aware imager is available. 
We used the flux-corrected image to obtain an updated sky model (again with {\tt PyBDSM}) for two subsequent rounds of self-calibration (the calibration strategy remaining unchanged). An example of the phase  self-calibration solutions obtained is given in Fig.~\ref{fig:LBAphasesol}. 

The final images (robust~=~$-0.1$) have noise levels of 10, 43, and 250~mJy~beam$^{-1}$ at 60, 30, and 20~MHz, respectively. The noise level at 63~MHz is about a factor of two lower than the deepest VLA 74~MHz images from \cite{2005ASPC..345..203L}. A lower resolution (robust~=~0.5 weighting) image at 63~MHz has a noise of 25~mJy~beam$^{-1}$, see Table~\ref{tab:lbaimages} for a summary of the resolution, bandwidth and sensitivity of the images.   
The thermal noise is about 2.5~mJy~beam$^{-1}$ at 63~MHz and 8~mJy~beam$^{-1}$ at 30~MHz. Ionospheric phase errors likely dominate the error budget, especially at 30 and 20~MHz. From the simulated point source grid, we estimate that errors coming from the usage of an imager that is unaware of the time-varying LOFAR beam, contribute about 2~mJy~beam$^{-1}$ to the noise at 63~MHz in the central part of the FOV. At the time of the A2256 observation the station calibration tables, which contain gain correction factors for the individual dipoles,  were still in the process of being refined. This could have resulted in a somewhat reduced station sensitivity.

\begin{table}
\begin{center}
\caption{LOFAR LBA image characteristics}
\begin{tabular}{lllll}
\hline
\hline
frequency & bandwidth &  obtained RMS & synthesized beam \\
      MHz     &      MHz                    & mJy~beam$^{-1}$    & arcsec        \\
\hline
63 (robust~-0.1) & 5.5      & 10   & $22\times26$     \\
63 (robust~0.5) & 5.5      & 25  &   $52\times62$ \\
30 (robust~-0.1)  & 3.7      &  43  & $58\times69$   \\
20 (robust~-0.1) & 3.7      &  250 & $108\times 116$\\
\hline
\hline
\end{tabular}
\label{tab:lbaimages}
\end{center}
\end{table}

The presence of residual ionospheric phase errors after calibration lead to a wider point spread function (PSF). We measure a FWHM of the PSF that is about 20\arcsec~larger than the synthesized beamwidth at 63~MHz in the central part of the FOV. This effect has also been seen for 74~MHz VLA observations \citep[e.g.,][]{2007AJ....134.1245C}.  The increase in the PSF FWHM varies between 20\arcsec~and 40\arcsec~across the FOV, according to the {\tt PyBDSM} PSF characterization module. We also compared the positions of the brightest sources in the FOV against the catalogue positions from NVSS and WENSS. We find the positional accuracies to be 5\arcsec~or better, depending on the SNR and location of the sources. At lower frequencies, the PSF FWHM is about 70\arcsec~and 180\arcsec~larger (in the central part of the FOV) than the synthesized beamwidth at 30 and 20~MHz, respectively. In this case the unresolved sources are elongated, distorted, and partly broken up into smaller components. For this reason,  we only use the 63 MHz maps for quantitative analysis in this paper, while the 30 and 20 MHz maps are used for qualitative analysis until a more complete ionospheric correction is made.

To obtain and check the overall flux scale we measured the integrated fluxes for five bright sources in the FOV:  4C~+79.16 , 8C~1654+785, 8C~1657+780, 8C~1708+779, and 8C~1728+783 (see Fig.~\ref{fig:LBApb}). For these sources we collected flux density measurements from the 1.4~GHz NVSS \citep{1998AJ....115.1693C}, 325~MHz WENSS  \citep{1997A&AS..124..259R}, 151~MHz 7C \citep{2007MNRAS.382.1639H}, 74~MHz VLSS \citep{2007AJ....134.1245C} and 38~MHz 8C \citep{1995MNRAS.274..447H} surveys.  
We fitted second order polynomials to these flux density measurements in $\log{(S)}-\log{(\nu)}$ space and compared the LOFAR flux density measurements at 63~MHz against the predicted fluxes from the polynomial fits. The median of the correction factors was used to tie the LOFAR images to the flux-scale from these surveys. The correction factor we found was modest, being 1.23 at 63~MHz. The spread in the individual correction factors is about 7\%, see Fig.~\ref{fig:LBAfluxratio}. 

From this we adopt an error in the relative LOFAR flux scale within the primary beam that is uncertain by 7--10\%.

\section{Results}
\begin{figure*}
   \begin{center}
       \includegraphics[angle = 0, trim =0.0cm 0.0cm 0.0cm 0.0cm,width=0.8\textwidth]{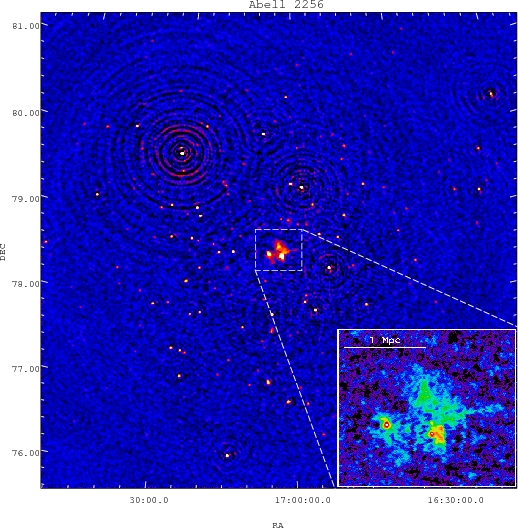}
     \end{center}
    \caption{Overview of the A2256 field at 61--67~MHz as observed with LOFAR. The image was not corrected for the primary beam attenuation and has a resolution of about  80\arcsec. Bottom right: Zoomed version of the 61--67~MHz image with a synthesized beam of $22\arcsec\times26\arcsec$.}
    \label{fig:lbaoverview}
\end{figure*}
\label{sec:results}

\subsection{Radio images}
\begin{figure*}
   \begin{center}
       \includegraphics[angle = 90, trim =0.0cm 0.0cm 0.0cm 0.0cm,width=0.49\textwidth]{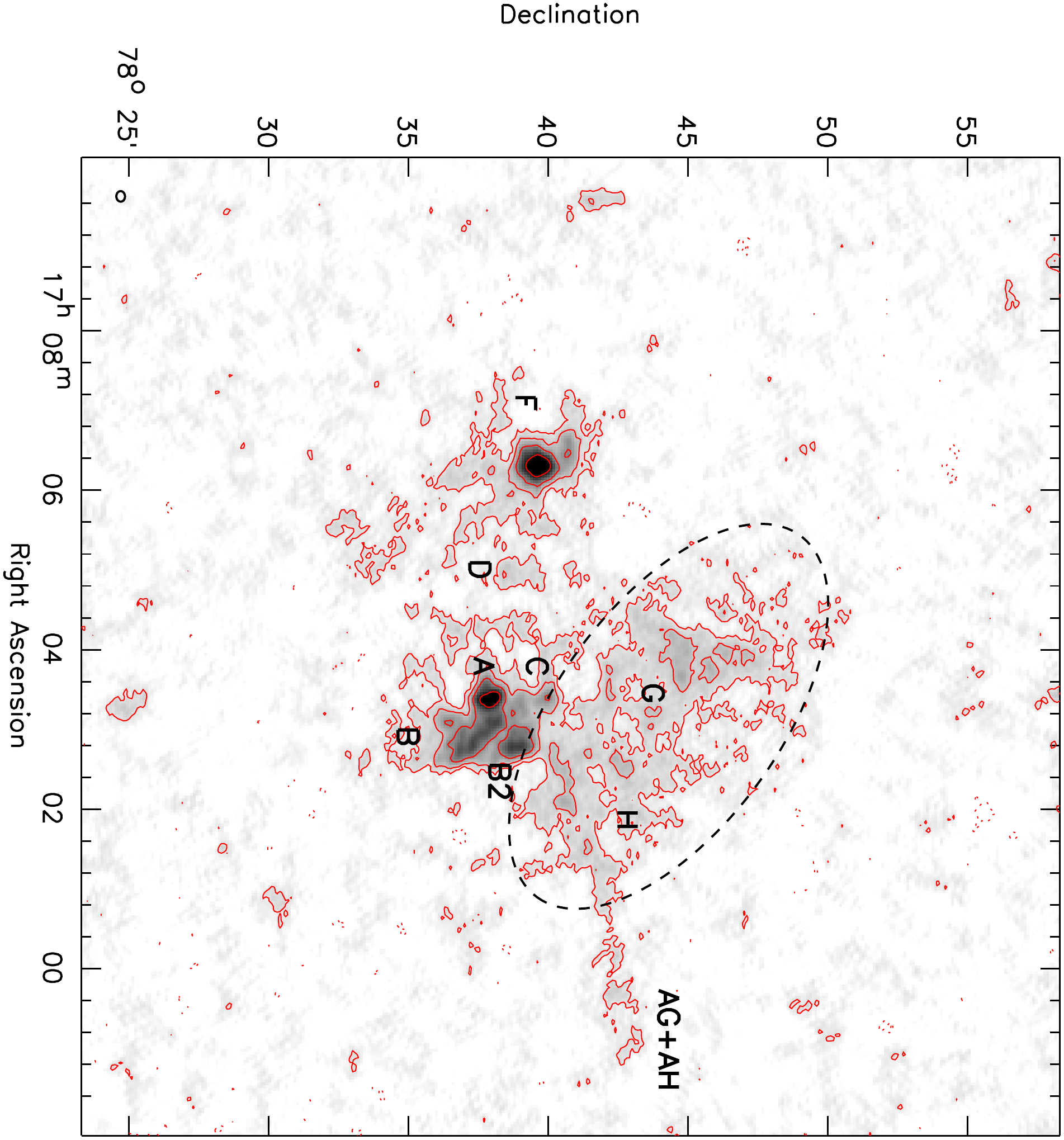}
       \includegraphics[angle = 90, trim =0.0cm 0.0cm 0.0cm 0.0cm,width=0.49\textwidth]{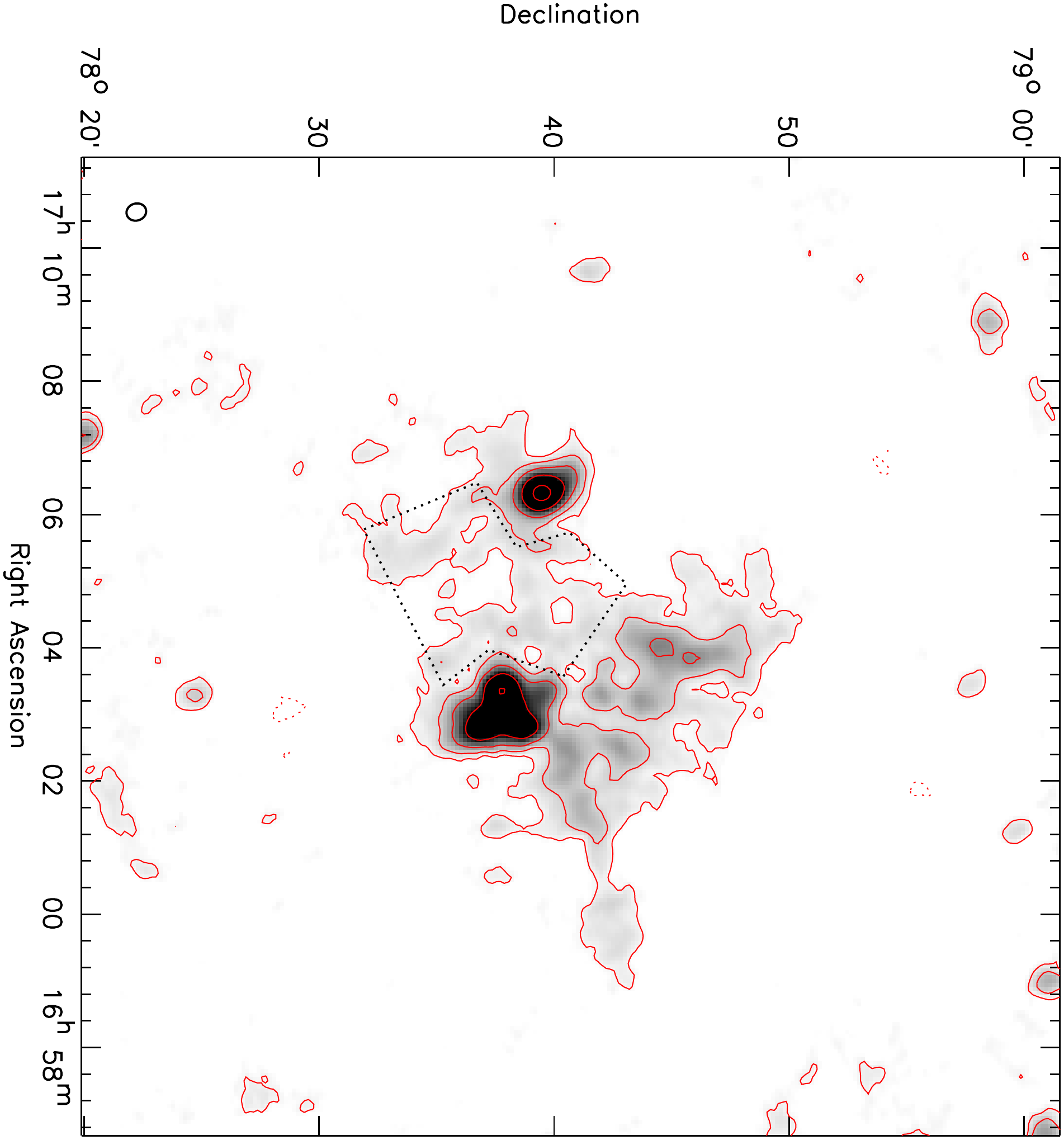}
     \end{center}
    \caption{A2256 61--67 MHz images represented in grey scale and contour form. The synthesized beams are shown in the bottom left corner. {\it Left}: High-resolution, $22\arcsec\times26\arcsec$, image made with robust~=~-0.1 weighting. The radio relic (source G and H) is indicated with the dashed ellipse. Contour levels are drawn at $[1, 2, 4, \ldots]~\times 30$~mJy~beam$^{-1}$.  Negative $3\sigma_{\mathrm{rms}}$ contours are shown by the dotted lines. {\it Right}: Low-resolution image, $52\arcsec\times62\arcsec$, image made with robust~=~0.5 weighting. The area for the halo spectral index measurement is indicated by the dotted lines. Contour levels are drawn at $[1, 2, 4, \ldots]~\times 75$~mJy~beam$^{-1}$.}
    \label{fig:lbaimages}
\end{figure*}

\begin{figure*}
   \begin{center}
       \includegraphics[angle = 90, trim =0.0cm 0.0cm 0.0cm 0.0cm,width=0.49\textwidth]{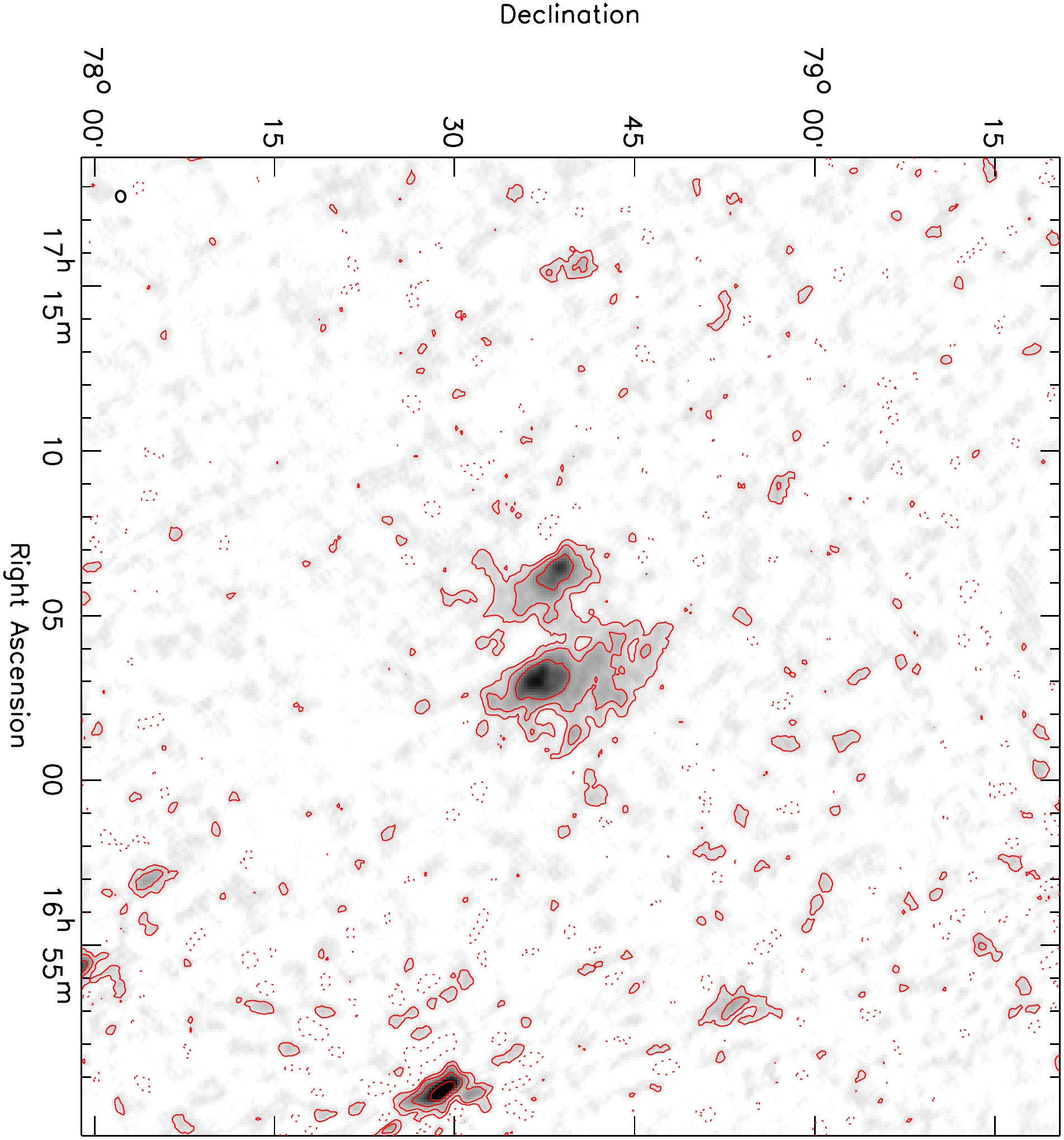}
       \includegraphics[angle = 90, trim =0.0cm 0.0cm 0.0cm 0.0cm,width=0.49\textwidth]{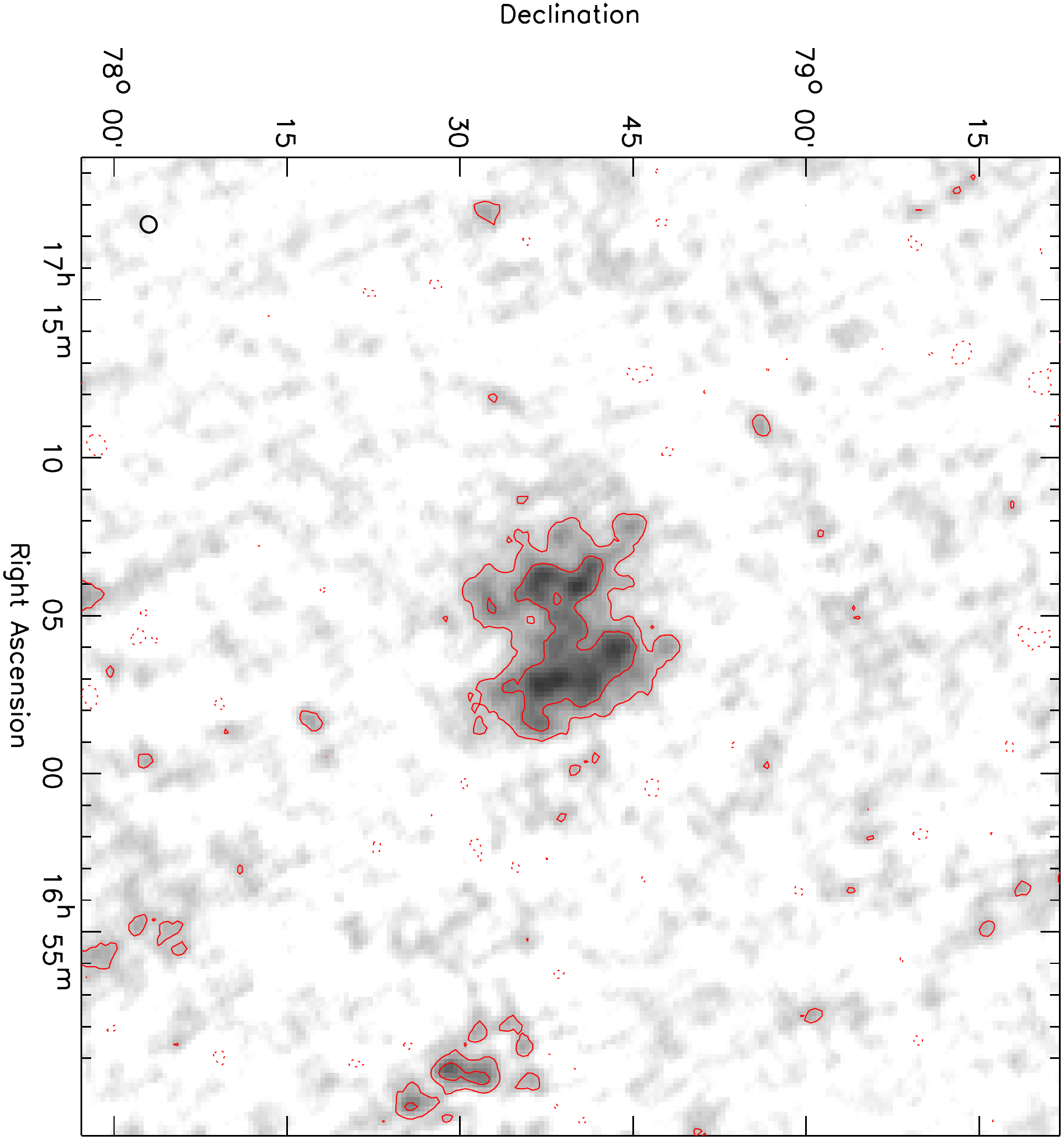}
     \end{center}
    \caption{A2256 30 and 20~MHz images represented in grey scale and contour form. The synthesized beams are shown in the bottom left corner. {\it Left}: 28--32~MHz image with a resolution of $58\arcsec\times69\arcsec$. Contour levels are drawn at $[1, 2, 4, \ldots]~\times 129$~mJy~beam$^{-1}$.  Negative $3\sigma_{\mathrm{rms}}$ contours are shown by the dotted lines. {\it Right}: 18--22~MHz image with a resolution of $108\arcsec\times116\arcsec$. Contour levels are drawn at $[1, 2, 4, \ldots]~\times 0.75$~Jy~beam$^{-1}$. Both images were made with robust~=~0.1 weighting. The bright compact source to the west around RA~16$^{h}$51$^{m}$ is 8C~1654+785.} 
    \label{fig:lba2030images}
\end{figure*}

An overview image of the field around A2256 is shown in Fig.~\ref{fig:lbaoverview}. The 63~MHz images are displayed in Fig.~\ref{fig:lbaimages} and the 20 and 30~MHz images in Fig.~\ref{fig:lba2030images}.  We labelled some of the known sources in the cluster following the scheme from \cite{1976A&A....52..107B, 1994ApJ...436..654R, 2009A&A...508.1269V}. The LOFAR 63~MHz image reveals some of the well-known tailed radio sources (A, B, F), the main relic (G and H), and part of the radio halo. A hint of the long and straight head-tail source C is also visible.

The main radio relic and halo are somewhat better visible in the lower resolution 63~MHz image (Fig.~\ref{fig:lbaimages}, right panel). Interestingly, the LOFAR image also reveals source AG+AH. This source has only been discovered recently \citep{2009A&A...508.1269V} and is not visible in the deep VLA 1.4~GHz observations \citep{2006AJ....131.2900C}, implying a steep spectrum. We do not detect the steep spectrum source AI, but this is expected since the integrated flux of this source is about a factor of two lower than AG+AH at 325~MHz. 
The 30~MHz image reveals source F and the combined emission from A and B. The relic is also detected. At 20~MHz the ionospheric phase distortions are quite severe, causing the relic to partly blend with source F and the A+B complex.

\begin{figure}
   \begin{center}
    \includegraphics[angle = 90, trim =0cm 0cm 0cm 0cm,width=0.5\textwidth]{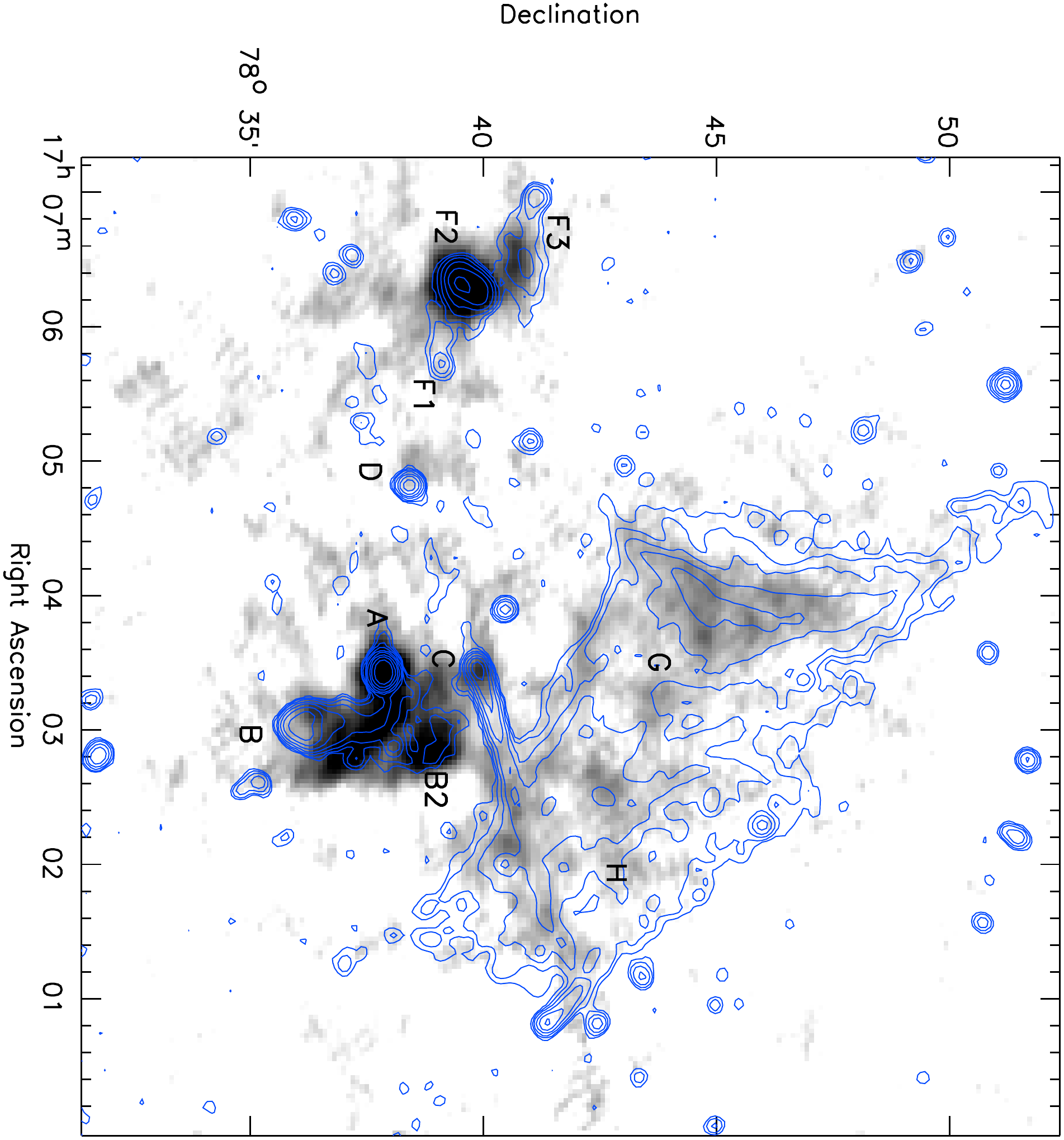}
     \end{center}
    \caption{LOFAR LBA 63~MHz image overlaid with VLA 1.4~GHz C-array contours. The contour levels are drawn at ${[1, 2, 4, 8,  \ldots]}~\times 90$~$\mu$Jy~beam$^{-1}$.}
    \label{fig:lfwsrt_lband}
\end{figure}

\begin{figure}
   \begin{center}
        \includegraphics[angle = 90, trim =0cm 0cm 0cm 0cm,width=0.49\textwidth]{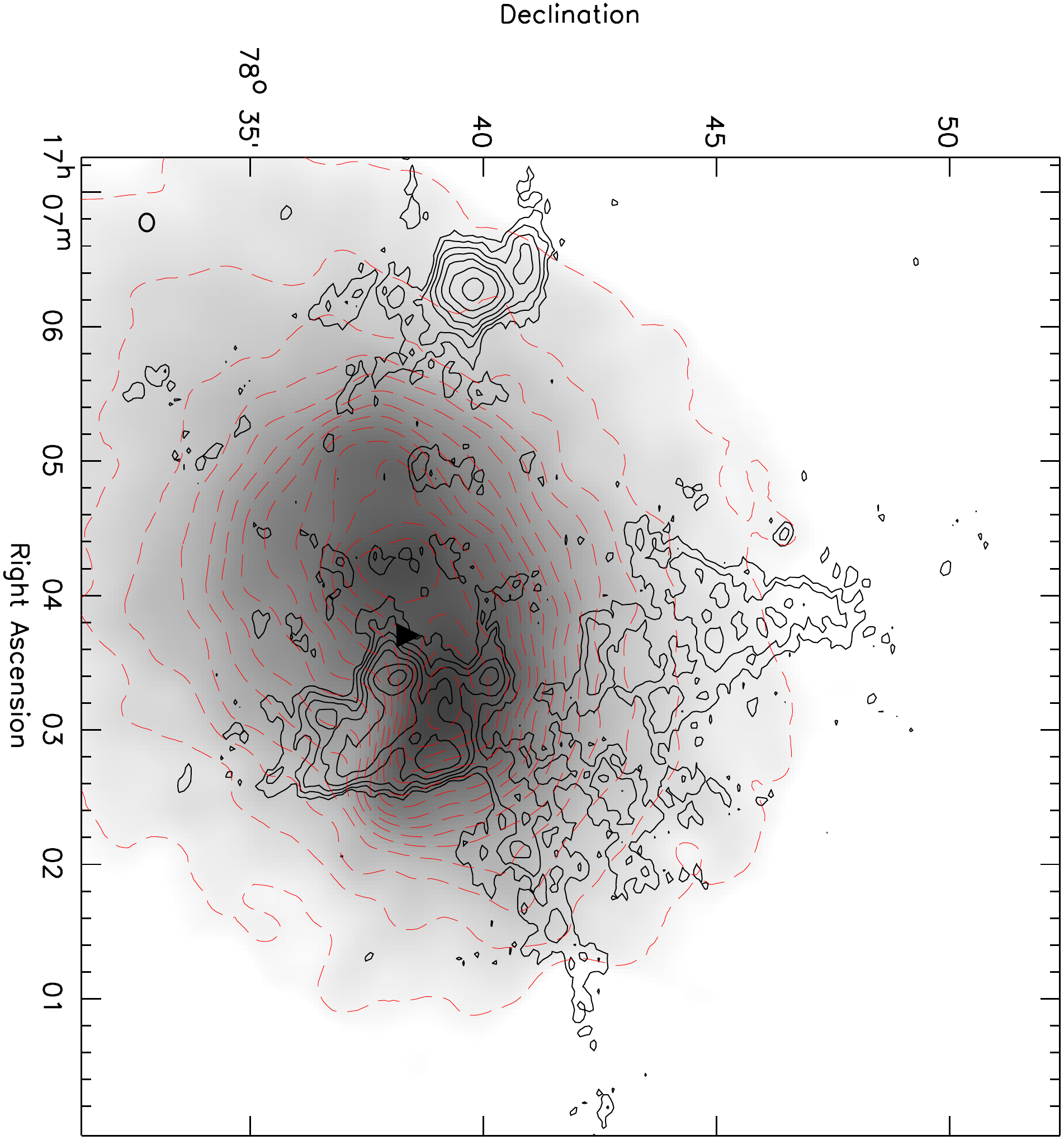}
             \end{center}
    \caption{Chandra 0.5--7~keV ACIS-I/S co-added image smoothed by a variable-width Gaussian, Fig.~1 from \cite{2002ApJ...565..867S}. Red dashed linearly spaced contours trace the X-ray  isophotes. The high-resolution 63~MHz LOFAR image is overlaid with black contours drawn at levels of ${[1, 2, 4, 8,  \ldots]}~\times 35$~mJy~beam$^{-1}$. The synthesized beam is shown in the bottom left corner. The triangle indicates the position of the main peak in the projected galaxy distribution derived by \cite{2007A&A...470...39R}.}
    \label{fig:LBAXray}
\end{figure}

An overlay of the 63~MHz image with a VLA 1.4~GHz image is shown in Fig.~\ref{fig:lfwsrt_lband}. For this we combined the 1369, 1413, 1513, and 1703 MHz VLA C-array images from \cite{2006AJ....131.2900C}. These images were convolved to a common resolution of $16\arcsec \times 16\arcsec$ and then combined, adopting a flux scaling according to a spectral index of $\alpha=-1$. It is interesting to note the differences between the VLA and LOFAR LBA image around the A+B complex. The VLA image peaks in brightness at the heads of the head-tail sources, while the LOFAR image mainly shows the tails.  The steep spectrum B2 region, noticed by \cite{intema_phd}, also clearly stands out. A Chandra X-ray overlay is shown in Fig.~\ref{fig:LBAXray}.  The X-ray emission in elongated in the NW-SE direction, with the relic located on the NW edge of the X-ray emission. The two main X-ray peaks are located between source D and A and just north of source A.

\subsection{Spectral index map and integrated fluxes}
\begin{figure*}
   \begin{center}
    \includegraphics[angle = 90, trim =0cm 0cm 0cm 0cm,width=0.49\textwidth]{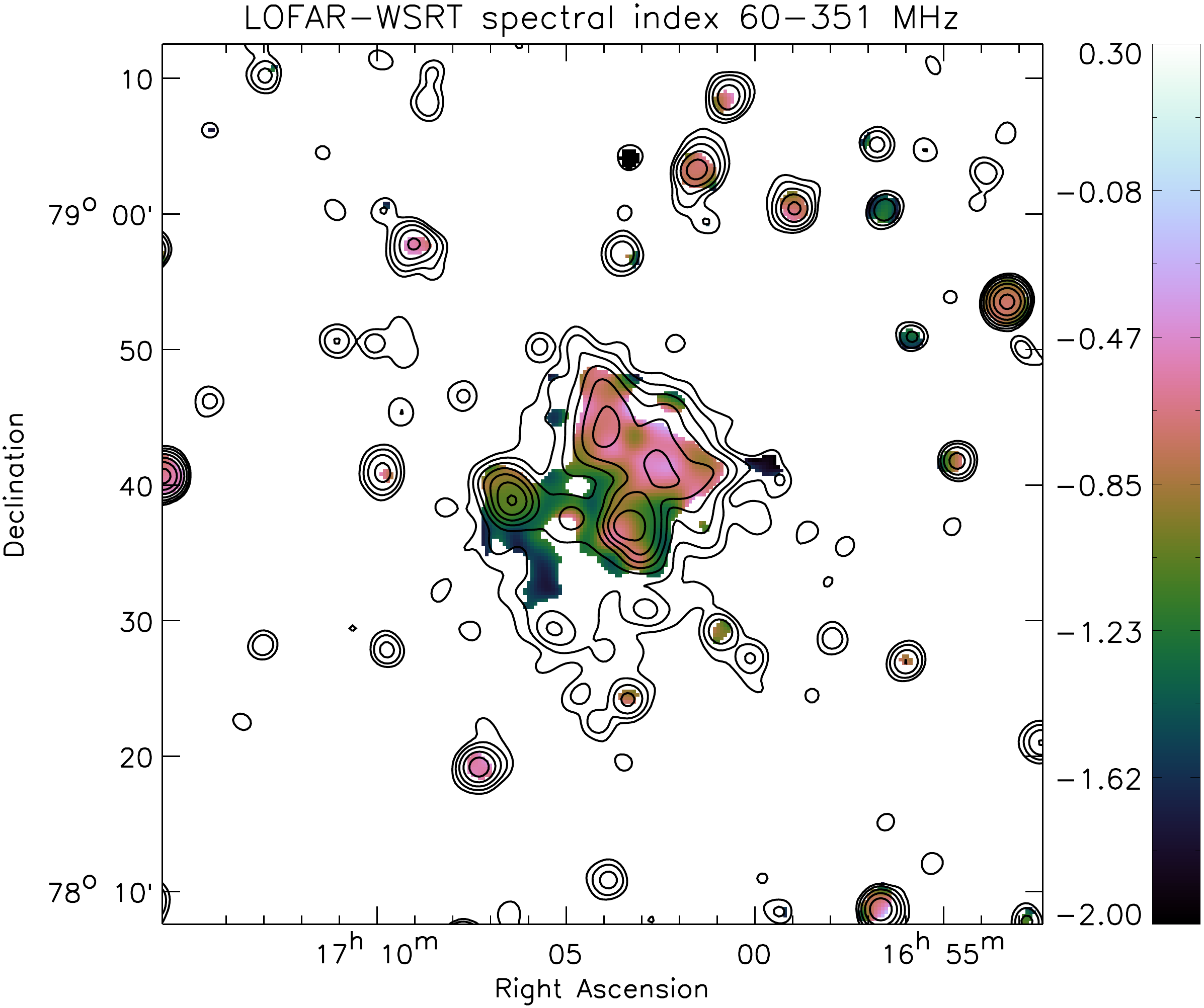}
    \includegraphics[angle = 90, trim =0cm 0cm 0cm 0cm,width=0.49\textwidth]{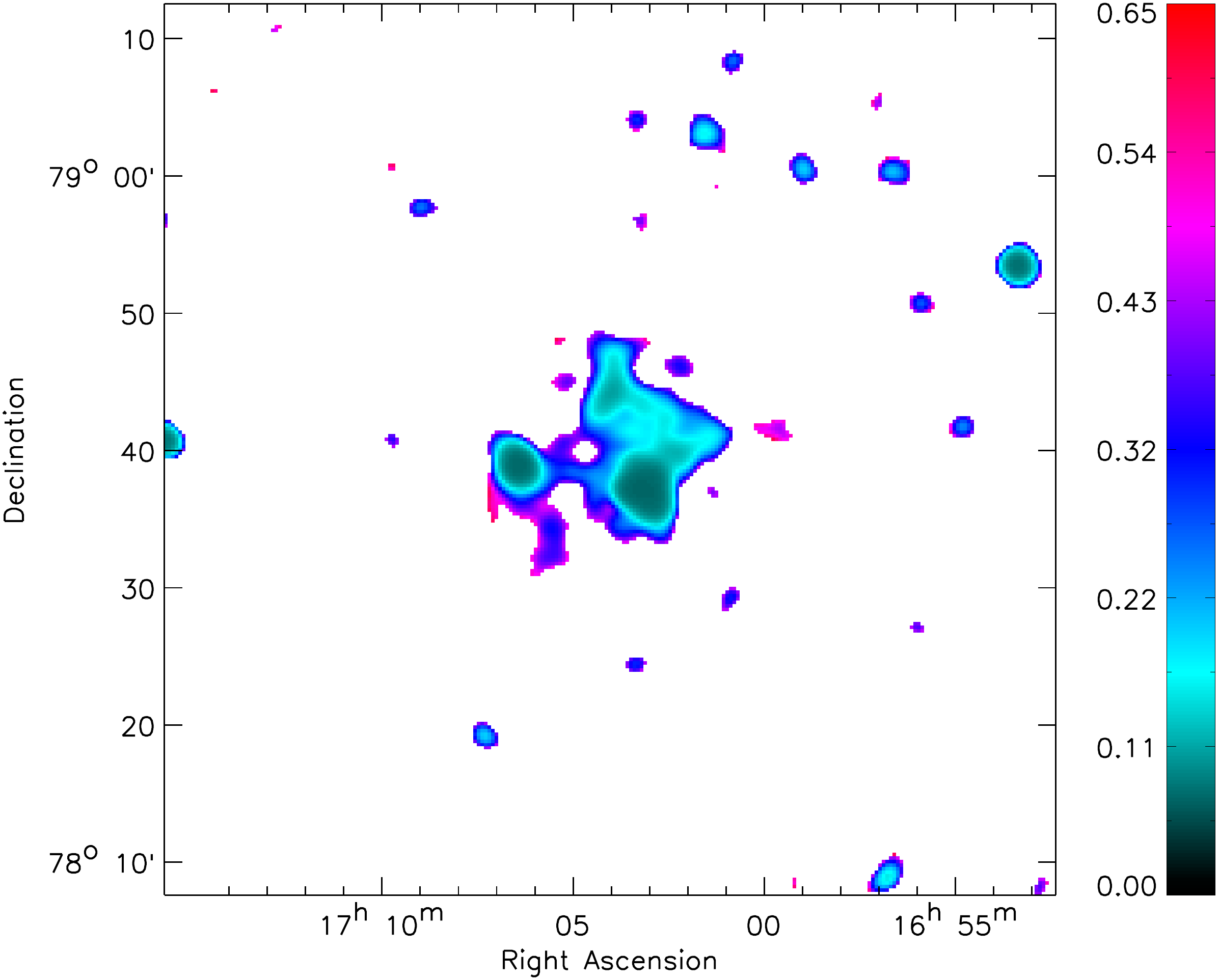}
     \end{center}
    \caption{{\it Left}: LOFAR LBA -- WSRT spectral index map between 63 and 351 MHz. Total intensity contours at 351 MHz are shown at levels of  ${[1, 2, 4, 8,  \ldots]}~\times 1.0$~mJy~beam$^{-1}$.  The 351~MHz image is taken from \cite{2008A&A...489...69B}. {\it Right}: Spectral index error map. The map is computed on the basis of the $\sigma_{\mathrm{rms}}$ values for the individual maps and the reported flux calibration uncertainty of 10\%.} 
    \label{fig:lfwsrt_spix}
\end{figure*}

We made a spectral index map between 63 and 351~MHz, making use of the WSRT 351~MHz map from \cite{2008A&A...489...69B}. 
 The WSRT map was made with uniform weighting of the visibilities and a Gaussian taper was applied in the uv-plane to create a circular  $67\arcsec$~beam. We also made a LOFAR image with uniform weighting and a Gaussian taper to match the WSRT resolution of $67\arcsec~\times 67\arcsec$. We applied an inner uv-range cut to the LOFAR data to match up the inner uv-range limit of the WSRT data. To increase the SNR per beam for the diffuse emission we convolved both images to 100\arcsec~resolution, pixels below $3\sigma_{\rm{rms}}$ were blanked. The LOFAR-WSRT spectral index map and the corresponding spectral index error map are shown in Fig.~\ref{fig:lfwsrt_spix}. 

The spectral index map reveals that the relic has a relatively flat spectral index of about $-0.8$, with variations of $0.3$ in $\alpha$ across the structure. For those parts of the radio halo where the spectral index can be measured, we find $\alpha$ to be in the range $-1.0$ to $-1.7$. For source F the spectral index is around $-1.0$.  For the combined emission from  AH and AG we find $\alpha$ to be between $-2.2$ and $-1.7$. 

\label{sec:spectralindex}

We  extracted the integrated fluxes for sources in the cluster from the 63~MHz image, see Table~\ref{tab:lbaflux}.
The integrated fluxes of the radio halo and relic (source G+H) are difficult to measure because they are partly blended with some of the complex head-tail radio sources in the cluster. To estimate their flux contribution, we used both the high and low resolution images (Fig.~\ref{fig:lbaimages}). From the high resolution image we measured the fluxes for source F, and the combined emission from A and B (A+B). Head-tail source C contributes a significant amount of flux in the region of the relic, as judged from the GMRT 153~MHz image from \cite{intema_phd}. In the 63~MHz image the source blends with the radio relic, making it impossible to obtain a reliable flux estimate. Using the 153~MHz GMRT image (giving a flux of $0.48 \pm 0.05$~Jy) and the reported 327~MHz VLA flux \citep[$0.247\pm 0.020$~Jy;][]{1994ApJ...436..654R} we find $\alpha =  -0.87\pm 0.17$ for source C. Extrapolating this to 63~MHz, we estimate a flux of 1.05~Jy. To measure the relic flux we summed the flux over the same region as indicated in Fig.~10  by \cite{2008A&A...489...69B}. After subtracting the flux contribution for source C, we obtain a relic flux of $5.6\pm0.8$~Jy. A power-law fit through the relic's flux  measurements at 1369 \citep{2006AJ....131.2900C}, 351 \citep{2008A&A...489...69B}, 153 \citep{intema_phd}, and 63~MHz gives a spectral index of $-0.81\pm 0.03$, see Fig.~\ref{fig:LBA_2256RH}. 
For the radio halo we find a flux of $6.6\pm1.3$~Jy from Fig.~\ref{fig:lbaimages} (right panel), integrating over the entire halo area as defined by the 351~MHz image.

\begin{table}
\begin{center}
\caption{63 MHz source fluxes and spectral indices computed between 63 and 153 MHz.}
\begin{tabular}{lllll}
\hline
\hline
Source & $S_{\rm{63 MHz}}$ & $\alpha_{63}^{153}$ \\
              &      Jy                                  \\
\hline
A+B&$5.1\pm 0.6 $  &$-1.1\pm 0.2$ \\  
F& $2.8\pm 0.3 $ & $-1.2 \pm 0.2$  \\
AG + AH  &$0.75 \pm 0.10$ & $-2.3 \pm 0.4$  \\ 
relic (G + H)& $5.6 \pm 0.8$ & $-0.81\pm 0.03^{a}$ \\ 
halo & $6.6\pm 1.3$ & $-1.5\pm 0.1^{b}$\\ 
\hline
\hline
\end{tabular}
\label{tab:lbaflux}
\end{center}
$^{a}$ from a polynomial fit in  $\log{(S)}-\log{(\nu)}$ space, see Fig.~\ref{fig:LBA_2256RH}\\
$^{b}$ between 63 and 351~MHz, $\alpha$ taken from Fig.~\ref{fig:LBA_2256RH}\\
\end{table}

\section{Discussion}
\label{sec:discussion}
The results on the radio spectra for source~F, AG+AH, the radio relic, and the radio halo are discussed in the following subsections.
\subsection{Source F}
Source F is known for its complex Z-shape morphology, see Fig.~\ref{fig:lfwsrt_lband}, and steep radio spectrum \citep[e.g.,][]{1979A&A....80..201B}. The source comprises three smaller structures: F1, F2, and F3. The brightest component F2, has a toroidal filamentary shape \citep{1994ApJ...436..654R, 2003AJ....125.2393M, intema_phd}. F3 has been classified as a head-tail radio source associated with galaxy~122 \citep[named by][]{1989ApJ...336...77F}. No optical counterparts have been found for F1 and F2,  and their origin is still being debated. \cite{1979A&A....80..201B} suggested that all three components are the tail of  galaxy 122, this scenario is also discussed by \cite{2008A&A...489...69B}. In another scenario, F2 could be the compressed fossil radio plasma from previous episodes of AGN activity \citep{2001A&A...366...26E, 2002MNRAS.331.1011E}.  This agrees with the observed toroidal shape \citep{2002MNRAS.331.1011E}. In this case, the fossil plasma could also have originated from galaxy 122.

\begin{figure}
   \begin{center}
    \includegraphics[angle = 90, trim =0cm 0cm 0cm 0cm,width=0.49\textwidth]{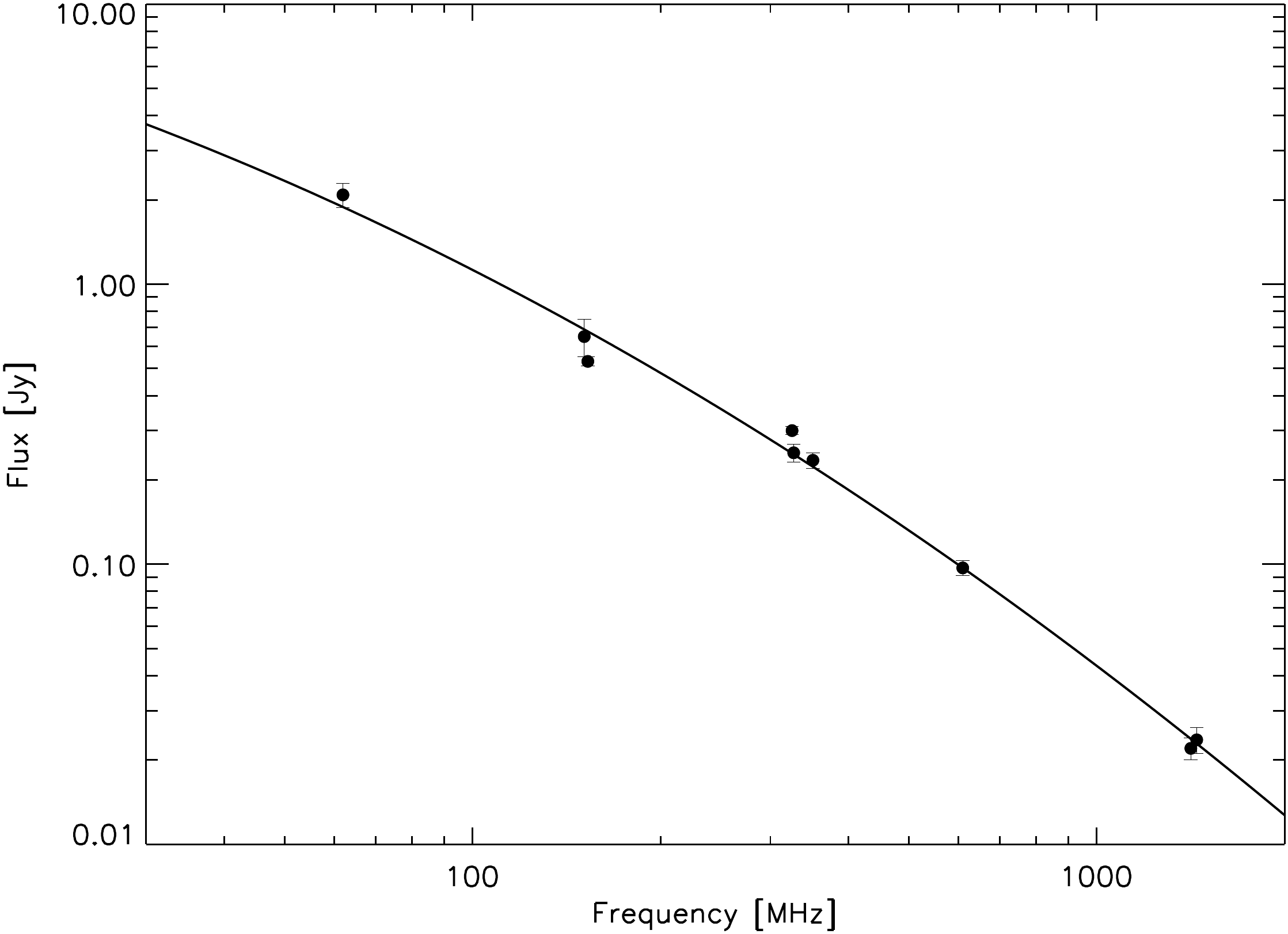}
             \end{center}
    \caption{Radio spectrum of F2. The fluxes were taken from \cite{2008A&A...489...69B}, except for the 63~MHz LOFAR flux.  The solid line is a second order polynomial fit through the flux density measurements in $\log{(S)}-\log{(\nu)}$ space. During the fitting procedure, we assumed 10\% uncertainty because the flux density measurements are not all consistent within their $1\sigma$ reported errors (plotted in the figure). This prevents the fit to be forced to go through a few measurements with small reported uncertainties.}
    \label{fig:LBAF2}
\end{figure}

We collected flux density measurements for F2 from the literature \citep[see][]{2008A&A...489...69B}, these are plotted in Fig.~\ref{fig:LBAF2}. The spectrum for F2 is clearly curved, with the spectrum flattening towards lower frequencies. From the fit, see Fig.~\ref{fig:LBAF2}, we find a spectral index of $-0.95$ between 63 and 153~MHz. Between 150 and 350~MHz \cite{2010ApJ...718..939K} reported $\alpha=-1.10 \pm 0.05$, but no flux density measurements were reported for F2. We find a spectral index of $-1.34$ between 150 and 350~MHz using the polynomial fit. The reason for the difference is unclear, but it should be noted that the measurements from the literature are not consistent within their reported uncertainties. Between 610 and 1400~MHz we obtain $\alpha=-1.67$, much steeper than at low-frequencies. In one scenario, \cite{2008A&A...489...69B} estimated a possible break frequency to be located at 26~MHz, assuming a constant magnetic field  of 7.3~$\mu$G and a spectral age of 0.2~Gyr. The 63~MHz flux density measurement  indicates the spectrum continues to flatten. However, high-resolution measurements below 60~MHz are needed to determine the possible underlying (``zero ageing'') power-law component. The origin of source F2 remains unclear, although it is likely the source is somehow related to the fossil radio plasma from previous phases of AGN activity given its brightness and steep radio spectrum.

\subsection{Source AG+AH}
The combined emission from source AG+AH is detected in the LOFAR 63~MHz image. With the non-detection of these sources in the deep 1.4~GHz image from \cite{2006AJ....131.2900C}, \cite{2009A&A...508.1269V} determined that $\alpha < -1.95$ between 325 and 1369~MHz. Between 140 and 351~MHz, using WSRT observations, the spectral index is $-2.05 \pm 0.14$ at a resolution of $175\arcsec$. At this resolution the flux of this feature partly blends with the relic emission. With the 153~MHz image from \cite{intema_phd} we find a flux of $95\pm10$~mJy for the combined emission from AG+AH. From the GMRT 325~MHz image a flux of $19 \pm 2$~mJy was reported \citep{2009A&A...508.1269V}. This gives a spectral index of $-2.1\pm0.2$ between these two frequencies, in agreement with the earlier reported result of $-2.05 \pm 0.14$. The LOFAR flux is $0.75 \pm 0.10$~Jy for this source. This results in a spectral index of $-2.3 \pm 0.4$ between 63 and 153~MHz. At low-frequencies the spectrum thus remains steep, although the uncertainty in the spectral index is too large to rule out a spectral turnover. 

We consider the possibility that this emission is related to the head-tail source C. Following \cite{1980ARA&A..18..165M} and \cite{2008A&A...489...69B} for a constant magnetic field and no adiabatic losses, the radiative lifetime/spectral age ($\tau$) is given as
\begin{equation}
\tau =  \frac{2.6 \times 10^{10}}{B^2 + B_{\rm{CMB}}^2  }   \sqrt{\frac{B}{(1+z)   \nu_{\rm{brk}}}}                  \mbox{ } \mbox { [yr]} \mbox{ ,}
\label{eq:lbatage}
\end{equation}
where $B$ is the magnetic field strength in $\mu$G, $B_{\rm{CMB}} \, [\mu\rm{G}]= 3.25(1+z)^2$  the equivalent magnetic field strength of the microwave background, and $\nu_{\rm{brk}}$ the break frequency  in MHz. The LOFAR 63~MHz flux density measurement indicates that $\nu_{\rm{brk}}$ is located $\lesssim 50$~MHz because the spectrum is still very steep between 63 and 153~MHz. The magnetic field strength is difficult to estimate as the spectral shape is poorly constrained. With a reasonable value of 10~$\mu$G we obtain a spectral age of 0.1~Gyr. This increases to 0.2~Gyr for $B=5$~$\mu$G. In all cases we assumed $\nu_{\rm{brk}}=50$~MHz. Spectral ages for different break frequencies and magnetic field values are listed in Table~\ref{tab:lbaagef}. \cite{1994ApJ...436..654R} estimated a velocity ($\rm{v}$) of $\sim2000$~km~s$^{-1}$ for the head-tail source. Then the separation between the ``head'' and AG+AH would be $\rm{v}\times \tau = 200$~kpc, for $B=10$~$\mu$G. AG+AH is located at a projected distance of about 800~kpc from the head of C. A lower magnetic field strength of $B=3.0$~$\mu$G would give a distance of 600~kpc. This is still lower than the projected distance of 800~kpc between the head of C and AG+AH. Therefore $\nu_{\rm{brk}}$ could be located at a lower frequency than 50~MHz (Table~\ref{tab:lbaagef}).  
 If the radio plasma from the tail (the AG+AH part) is compressed by the merger shock wave, the radiative age of the source can also be older \citep[about 0.5~Gyr or more, e.g.,][]{2009ApJ...698L.163D, 2012ApJ...744...46K}. In this case Eq.~\ref{eq:lbatage} does not apply.  This makes it easier to explain the distance of at least 800~kpc, even more since this distance is a lower limit because of unknown projection effects. Given all these uncertainties, source AG+AH could indeed be an old part of the tail of source C.

\begin{table}
\setlength{\tabcolsep}{0.1cm}
\begin{center}
\caption{Spectral age of the source AG+AH in Gyr for different break frequencies and magnetic field strengths. Values in boldface are sufficient to explain the distance of at least 800~kpc between the head of the source C and AG+AH, in the scenario that AG+AH is a part of the tail of C.} 
\begin{tabular}{l | llll}
\hline
\hline
& $B=3$~$\mu$G & $B=5$~$\mu$G & $B=10$~$\mu$G & $B=20$~$\mu$G \\
\hline
$\nu_{\rm{brk}} = 50$~MHz & 0.28 & 0.21 & 0.10 & 0.039\\
$\nu_{\rm{brk}} = 20$~MHz & \bf{0.44} & 0.33 & 0.16 &0.061\\
$\nu_{\rm{brk}} = 5$~MHz & \bf{0.88}   &  \bf{0.66} & 0.32 & 0.12\\
$\nu_{\rm{brk}} = 1$~MHz & \bf{2.0}      & \bf{1.5}    & \bf{0.71} & 0.27\\
\hline
\hline
\end{tabular}
\label{tab:lbaagef}
\end{center}
\end{table}

\subsection{Radio relic}
Radio gischt relics are proposed to trace particle acceleration (or re-acceleration of seed particles) at shocks. 
In the linear test particle regime and assuming that particles can diffuse
across the shock discontinuity (DSA), the spectrum of the re-accelerated
particles is given by \citep[e.g.,][]{1987PhR...154....1B}:
\begin{equation}
N(p)= (-\delta_{inj}+2) p^{\delta_{\rm{inj}}} \int_{p_{\rm{min}}}^{p_{\rm{max}}} N_s(p) p^{-\delta_{\rm{inj}}-1} dp
\end{equation}
where $N_s(p)$ is the spectrum of seed particles and  
\begin{equation}
\delta_{\rm{inj}} = -2 {{{\cal M}^2 +1}\over{{\cal M}^2 -1}}  \mbox{ ,}
\label{eq:lbainj-mach}
\end{equation}
${\cal M}$ is the Mach number of the shock.
The corresponding synchrotron spectral index, not including energy losses, is  
$\alpha_{\rm{inj}} =  (\delta_{\rm{inj}} -1)/2$
(with $F(\nu) \propto \nu^{\alpha_{\rm{inj}}}$).

If the properties of the shock remain unchanged (i.e., ``stationary conditions''), the relic has been present in the ICM longer than the electron cooling time and the cooling time is shorter than the diffusion time, the integrated radio spectrum will be a power-law, with a spectral index that is about 0.5 units steeper than $\alpha_{\rm{inj}}$ \citep{1962SvA.....6..317K}.  The radio spectrum of the A2256 relic has a power-law shape over the observed frequency range, with a spectral index of $-0.81\pm 0.03$. This would imply $\alpha_{\rm{inj}} \sim -0.3$ for a simple shock model. 
The flattest possible injection spectral index is $-0.5$ from DSA, which is not consistent with $\alpha_{\rm{inj}} \sim -0.3$. Therefore this directly implies that  stationary conditions for the shock do not apply and/or the acceleration process is more complicated\footnote{The electron cooling time can also not be longer than the diffusion time. At 1.4~GHz, where the relic has about the same size as at 63 MHz, the electron cooling time is orders magnitude smaller than the diffusion time \citep[e.g.,][]{1977ApJ...212....1J, 2002NewA....7..249B}.}. 
It could suggest that the relic brightened very recently ($<0.1$~Gyr or so) due to an increase in the Mach number while the shock front propagated outward, since the efficiency of DSA is believed to increase strongly with Mach number \citep[e.g.,][at least in this regime with $\mathcal{M} \lesssim 4$]{2007MNRAS.375...77H}. Recently injected relativistic electrons would thus have a higher density and hence a higher brightness. Therefore, the average spectral slope below the flattest value expected from standard DSA may reflect the strengthening of the shock front while moving outwards. In this case no equilibrium has yet been reached between the electron cooling and injection within the observed frequency range below 1369~MHz, but  one expects spectral steepening at higher frequencies. \cite{2006AJ....131.2900C} indeed reported a steeper spectral index of $-1.2$ between 1369 and 1703~MHz for the relic. For a typical magnetic field strength of 2~$\mu$G for the relic \citep{2006AJ....131.2900C} and $\nu_{\rm{brk}}=1400$~MHz, the spectral age is about 0.05~Gyr. 

\cite{2002AJ....123.2261B} proposed that A2256 is undergoing a triple merger event. One between a subcluster and the primary cluster and one between the primary cluster and subcluster (or the primary cluster only) with a smaller galaxy group. The relative velocity between the subcluster and primary cluster is estimated to be $\sim2000$~km~s$^{-1}$ and they are near the time of the first close passage of the subcluster and primary cluster centers. The group plunges down from the north. The merger between the primary cluster and subcluster has a mass ratio of about 3 and the merger between the primary cluster (+subcluster) and group has a mass ratio of about 10. 
\cite{2003AJ....125.2393M} argued that the ``primary cluster--galaxy group'' merger is responsible for the radio relic and that the merger event is viewed 0.3~Gyr after the core passage. 
From the radio spectra alone it is not possible to disentangle the merger scenario, but usually the strongest shocks form after core passage \citep{2012MNRAS.tmp.2652V} and the flat  integrated radio spectrum implies that the relic only recently brightened. Further flux density measurements above 2~GHz are needed to better constrain the high-frequency end of the spectrum and confirm the radio spectrum steepens here.

No shock has been found so far in X-ray observations \citep[e.g.,][]{2008A&A...479..307B}. Although, given the large extent of the relic in both the NS and EW directions we are probably not viewing the relic close to edge-on, making it more difficult to detect a shock. For an edge-on shock/relic one would expect a much larger ratio between the largest physical extent and the relic width \citep[e.g.,][]{2010Sci...330..347V}. The relic measures about 1 by 0.5~Mpc. We can estimate the viewing angle if we assume the relic traces a planar shock located in the xy-plane and has a similar extent in both the x and y directions and a negligible extent in the shock downstream region (compared to the observed width). With the ratio of the largest physical size to the relic width this gives a viewing angle of about 30\degr~from edge-on. This is consistent with the estimate from \cite{1998A&A...332..395E}, based on the polarization fraction, which indicates the relic is seen under an angle of less than $\sim 50\degr$ from edge-on. The relic is located at a project distance of $\sim 400$~kpc from the optical center of the cluster (Fig.~\ref{fig:LBAXray}). If the relic is seen under a viewing angle of 30\degr~from edge-on it is located at a true distance of $\sim 0.5$~Mpc from the cluster center. 
This is about a factor of two closer to the cluster center than most of the double radio relic clusters \citep[e.g.,][]{1997MNRAS.290..577R, 2006Sci...314..791B, 2009A&A...494..429B, 2011A&A...528A..38V}. Together with the flat radio spectrum this is consistent with the fact  that the relic in A2256 is seen at a relatively early stage in the merger, approximately half of the time after core passage compared to  some of the double radio relic clusters. 

Due to the large size of the relic it is unlikely we are seeing fossil radio plasma compressed by a shock wave, since radiative energy losses during the time it takes to compress a several hundred kiloparsec sized radio ghost would remove most of the electrons responsible for the observable radio emission \citep{2006AJ....131.2900C}. In addition, we would expect steep curved radio spectra due to synchrotron and IC losses.

The unusual flat spectrum of the relic may also suggest a more
complex situation. 
A synchrotron spectral index $\alpha > \alpha_{\rm{inj}} -1/2$ is expected
if electrons accelerated at the passage of the shock, with spectrum
$N(p) \propto p^{-\delta_{\rm{inj}}}$, are also
re-accelerated by some other mechanism in the region downstream
of the shock.
Under particular conditions a flatter spectrum is also expected 
if the accelerated electrons cool/age in a inhomogeneous 
downstream region.
Particle re-acceleration mechanisms downstream of large-scale shocks
in galaxy clusters, possibly connected with the
turbulence that could be generated by the shock passage,  
is suggested by the existence of bridges of faint radio emission
that connect relics and halos in several systems \citep[][and references therein]{2010arXiv1010.3660M}. 
In the case of A2256, projection effects may mix together 
the relic and bridge emission.

 \subsection{Radio halo}

\begin{figure}
   \begin{center}
    \includegraphics[angle = 90, trim =0cm 0cm 0cm 0cm,width=0.49\textwidth]{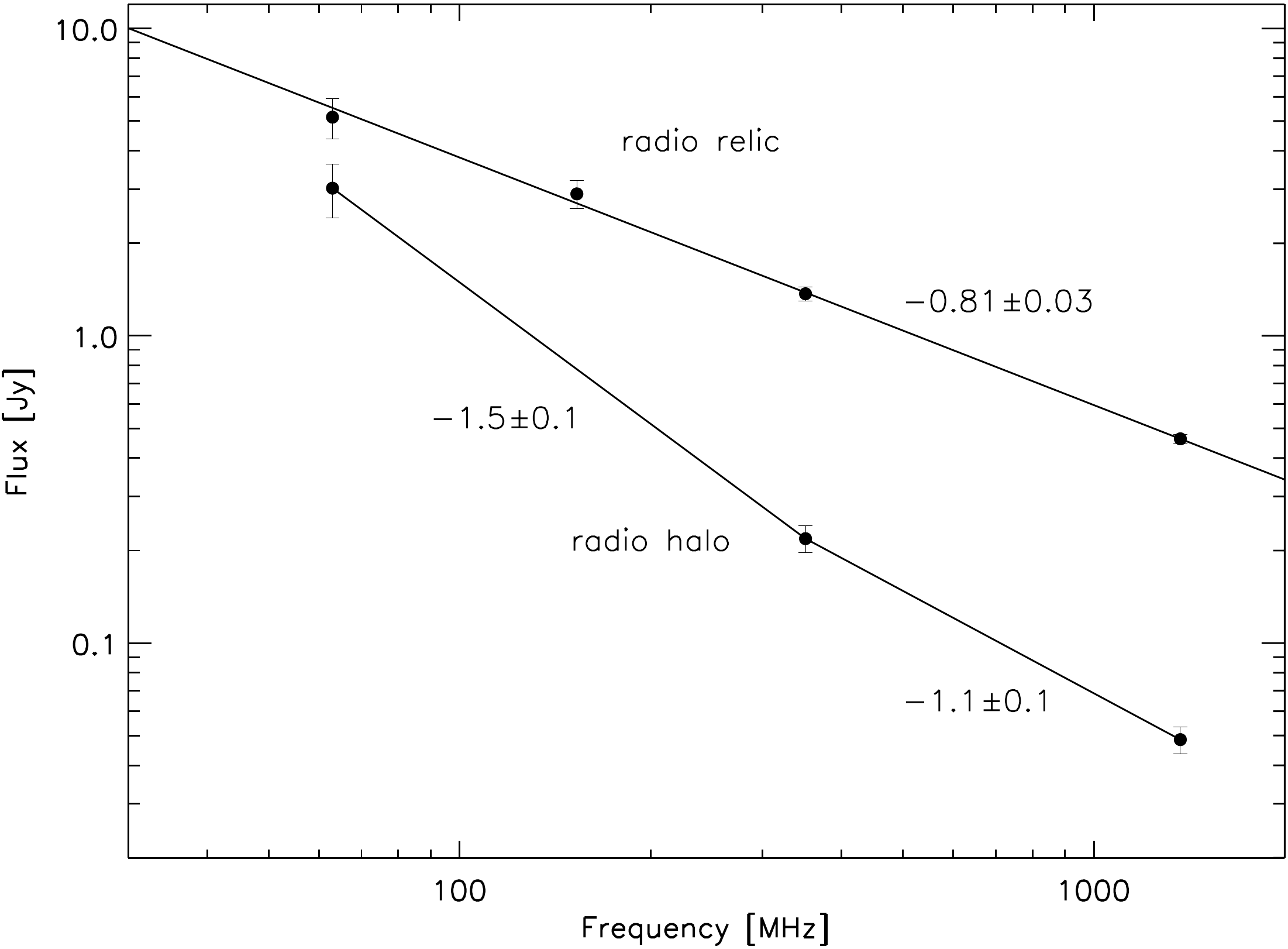}
     \end{center}
    \caption{Radio halo and relic spectrum. Flux density measurements from this work, \cite{2006AJ....131.2900C}, \cite{2008A&A...489...69B}, and \cite{intema_phd} were included. The solid straight line is a power-law fit to the relic fluxes. The radio halo spectrum is from flux density measurements summed over the region indicated in Fig.~\ref{fig:lbaimages}.}
    \label{fig:LBA_2256RH}
\end{figure}

The spectral index of the radio halo was measured by summing the flux in the region around source D, using the imaged displayed in Fig.~\ref{fig:lbaimages} (right panel). In this region, the halo is detected at the 1--3$\sigma$ level per beam in the LOFAR 63~MHz image. We summed the flux over the same region in the 351~MHz WSRT and 1369~MHz VLA D-array \citep{2006AJ....131.2900C} images. The resulting radio spectrum is shown in Fig.~\ref{fig:LBA_2256RH}. The 63--351~MHz spectral index is $-1.5 \pm 0.1$ and the 351--1369~MHz spectral index is $-1.1\pm0.1$. The low-frequency radio halo spectrum is steeper than at high-frequencies, although we only measured the spectrum in the area indicated in Fig.~\ref{fig:lbaimages}.

LOFAR observations at 63 MHz clearly show a significant
upturn of the spectrum of the radio halo at lower frequencies.
This confirms and strengthens previous observational claims
\cite{2010ApJ...718..939K} and provides totally new 
information for the interpretation of the origin of the halo.
With only three data points it is premature to attempt a detailed
modeling, this is matter for future papers when
more data will be available. 
The complex spectrum may simply result from a superposition of two (or more) components, as previously suggested by \cite{2010ApJ...718..939K}. Specifically, the emission of the relic could be projected on the halo emission causing a flattening of the spectrum at higher frequencies. We also note that recent modeling of turbulent re-acceleration 
of relativistic protons and their secondaries in the ICM predict a flattening of the synchrotron spectrum of radio halos
at higher frequencies \citep{2011MNRAS.410..127B}. In these models
the flattening marks the transition between the spectral component due to
turbulent re-acceleration and the underlying spectrum generated by the continuous
injection of secondary electrons; the spectral shape of A2256 would constrain the turbulent acceleration time-scale $\simeq 0.5$ Gyr.
Observations however may reveal situations more
complex than those considered in the presently available models.
Different populations of emitting electrons may coexist in the volume of
the radio halo in the case they originate from multiple acceleration 
mechanisms, or the efficiency of their acceleration changes with space
and time in the emitting volume. 
The latter hypothesis may reflect the fact that magnetic 
turbulence in a Mpc$^3$ region of the ICM is not homogeneous.
New observations with LOFAR at 150~MHz and complementary observations at high
frequencies are needed to determine the shape of the spectrum of the halo 
and to better constrain its origin.

\section{Conclusions}
We have presented initial results from LOFAR LBA observations between 18 and 67 MHz of the cluster Abell~2256. We focussed mainly on the 63~MHz map because at lower frequencies ionospheric phase distortions were severe. At 63~MHz we detect both the radio halo and main radio relic. The radio spectrum for the relic is consistent with a power-law, with $\alpha =-0.81\pm 0.03$. The integrated radio spectrum of the relic is quite flat, which could mean that the relic has only recently brightened due to an increase in the Mach number, within the last $\sim0.1$~Gyr. Alternatively, particles are re-accelerated by some mechanism in the downstream region of the shock.  
For the radio halo we find low-frequency spectral steepening, that was previously also reported by \cite{2010ApJ...718..939K}. 
Additional flux density measurements are needed to better determine the low-frequency spectrum of the halo and the mechanisms responsible for the acceleration of the emitting particles.

We detected a recently discovered steep spectrum source in the west of the cluster center, located roughly at the end of the previously known head-tail source C. For this source we find an extremely steep spectral index of $-2.3\pm0.4$ between 63 and 153~MHz. This steep spectrum source could be an older part of the tail of source C. For the source F2, we find that the spectral index flattens to $-0.95$ between 63 and 153~MHz. The origin of the source remains unclear.

In the future we plan to extend our investigation to lower frequencies. For this, ionospheric calibration schemes will be important to retain enough spatial resolution to separate the contribution from the various complex sources in the cluster.

\label{sec:conclusion}

\begin{acknowledgements}
We thank the anonymous referee for useful comments. LOFAR, the Low Frequency Array designed and constructed by ASTRON, has facilities in several countries, that are owned by various parties (each with their own funding sources), and that are collectively operated by the International LOFAR Telescope (ILT) foundation under a joint scientific policy.
We thank Ming Sun for providing Chandra X-ray image.  
We have used the ``cubehelix'' color scheme from \cite{2011BASI...39..289G}. 
RJvW acknowledges funding from the Royal Netherlands Academy of Arts and Sciences. MB, AB and MH acknowledge support by the Deutsche Forschungsgemeinschaft under grant FOR1254. GB and RC acknowledge partial support through PRIN-INAF 2009 and
ASI-INAF I/009/10/0. CF and GM acknowledge financial support by the ``Agence Nationale de la Recherche'' through grant ANR-09-JCJC-0001-01. Basic research in radio astronomy at the Naval Research Laboratory is supported by 6.1 Base funding. The Open University is incorporated by Royal Charter (RC 000391), an exempt charity in England \& Wales and a charity registered in Scotland (SC 038302). HTI is Jansky Fellow of the National Radio Astronomy Observatory. OW is supported by the `Deutsche Forschungsgemeinschaft' (Emmy-Noether Grant WU 588/1-1) and by the European Commission (European Reintegration Grant PERG02-GA-2007-224897 `WIDEMAPÕ).

\end{acknowledgements}

\bibliographystyle{aa}
\bibliography{ref_filaments_a2256}

\end{document}